\documentclass[12pt]{iopart}
\usepackage[utf8]{inputenc}
\usepackage[T1]{fontenc}
\usepackage{hyperref}
\usepackage{graphicx}
\usepackage[pdftex]{xcolor}

\usepackage{stix}
\usepackage{natbib}
\usepackage{xspace}
\usepackage{upgreek}

\newcommand{\phiMinus}{\varphi_-}%
\newcommand{\fPlus}{f_+}%
\newcommand{\uTilde}{\varphi}
\newcommand{\vTilde}{f}

\newcommand{\tfrac}[2]{\textstyle\frac{#1}{#2}}

\hypersetup{colorlinks=true,linkcolor=black,citecolor=black,urlcolor=blue,hyperfootnotes=true}

\newcommand{\bibcommenthead}{} % command used in bibliography
\begin{document} % here for iopart
\sloppy

%%%%%%%%%%%%%%%%%%%%%%%%%%%%%%%%%Title%%%%%%%%%%%%%%%%%%%%%%%%%%%%%%%%%
\title{Reflection phase shifts of bouncing Bogoliubov waves}
\author{%
Carsten Henkel%
\footnote[8]{henkel@uni-potsdam.de}
%\footnote[1]{1}\footnote[2]{2}\footnote[3]{3}\footnote[3]{3}
%\footnote[4]{4}\footnote[6]{6}\footnote[7]{7}\footnote[8]{8}
%\footnote[9]{9}
}
\address{University of Potsdam, Institute of Physics and Astronomy, 
Karl-Liebknecht-Str. 24/25, 14476 Potsdam, Germany}

%%%%%%%%%%%%%%%%%%%%%%Abstract%%%%%%%%%%%%%%%%%%%%%%%
\begin{abstract}
The Bogoliubov--de Gennes equations are solved for an inhomogeneous
condensate in the vicinity of a turning point, addressing the full
continuous spectrum.
A basis change in the space of the two Bogoliubov “particle” and
“hole” amplitudes is introduced 
that decouples them approximately.
We find a spatially extended mode 
that governs mainly excitations in the condensate phase,
while another mode is localised to regions with density gradients.
An analytical and numerical discussion of the phase shift is provided
that incident matter waves suffer upon reflection at the turning point,
forming standing waves.
As an application, we compute eigenfrequencies in a gravitational trap,
without recourse to the local density approximation.
The non-condensate density at finite temperature and its quantum depletion
are discussed in a companion paper.
\end{abstract}
%%%%%%%%%%%%%%%%%%%%%%%%%%%%%%%%%%%%%%%%%%%%%%%%%%%%%

%\pacs{}

%\submitto{J. Phys. B}
%let's see: two or three slices
%-- short paper with phase shift (is really very small and
%everything conspires to make it small!)
%-- longer paper for harmonic trap and mode shape near
%boundary
% -- published paper on 
%border-localised density spectra is already long (incl
%technical appendix (pedagogical) with Wronskians)

%\noindent
% µm and $\mu{\rm m}$ 
%2024 May 08 -- focus on the “other” potentials, hard-wall like or exponential
%(or even half-harmonic) and generalise the theory to these settings.

The boundary of a many-body system provides the arena to 
archetypal interference phenomena familiar from water waves.
On the microscopic scale, the surface of a dense system is often 
realised as a smooth cross-over between distinct phases. 
Similar to critical fluctuations in a phase transition, fluctuations 
are locally enhanced, as the stabilising effect of interactions 
in the high-density region is reduced. 
In this paper, we provide a detailed analysis of a degenerate bosonic gas
and its boundary,
as it may be realised in the Bose-Einstein condensed phase of ultracold 
atomic vapours \citep{PitaevskiiStringari}.
We consider a quasi-one-dimensional setting where the high-density phase 
is characterised by a reduction of density fluctuations, 
while thermal phase fluctuations destroy long-range order
\citep{ProukakisBook}.
The focus will be on the boundary region where due to a rise
in the trapping potential, the condensate density drops
so that a transition zone emerges that interfaces 
a dilute, quasi-ideal Bose gas and a dense degenerate gas.
Since there is no prediction for the equation of state
that is valid across this transition \citep{Kerr24},
little theoretical work
beyond the local-density approximation has been reported
when it comes to spatial density profiles and correlation functions
at finite temperature \citep{Oehberg97, Fedichev98, Fetter98a, 
Proukakis98b, Rusch99, Kheruntsyan05}.

The shape of the condensate wave function trapped in a potential
has been studied in detail, but with an understandable focus on 
its central, high-density region.
Its tails towards tunnelling regions below the confining potential
can be accurately described by a universal mathematical
function, the Painlevé II transcendent \citep{DLMF}, provided
the trapping potential can be linearised around its crossing 
with the chemical potential (i.e., the generalised eigenvalue of the
nonlinear Gross-Pitaevskii equation).
Our aim is to provide a physics-motivated analysis
of the elementary excitations around this solution, 
also known as Bogoliubov modes. 
In the mathematical literature, matching formulas for the Painlevé~II
function that provide its asymptotic form to the left and the right
of the turning point are well known \citep{Ablowitz77}, and these
also cover the case of an oscillatory solution. 
In physical problems, a particular solution of the Gross-Pitaevskii equation
is singled out by requiring that it follows the Thomas-Fermi asymptote
on the high-density side, given by
Eq.\,(\ref{eq:Thomas-Fermi-density}) below.
This particular case represents in a sense the separatrix 
between oscillatory and diverging solutions \citep{Hastings80}, 
at least for the linear potential.
This fact may play a role when it comes to linear perturbations 
around the condensate, since a naïve integration of the
differential equations from low to high densities leads to a divergence,
due to exponentially increasing solutions. 
We emphasise that this happens outside the tunnelling regime and
the solutions should behave as oscillatory functions.
The divergence does not show up for trapped systems where the 
oscillations cover only a finite range between two turning points
where the potential is below the chemical potential.

This paper builds on and develops further numerically stable techniques 
of earlier work for the continuous spectrum \citep{Diallo15a}.
The focus is on analysing the phase shift of the oscillatory Bogoliubov modes.
In scattering theory, such phase shifts provide essential information
about the potential that is accessible from the far-field asymptotics of
scattering solutions. 
They determine, for example, cross sections or the lifetime
of resonances \citep{MessiahI, Berry72}, 
and in the case of optics, an effective displacement of a reflected beam
similar to the Goos-H\"anchen shift \citep{Goos47, Henkel94a}.
We find that generic shapes of the potential in the vicinity of the turning
point lead to qualitatively different phase shifts when the energy of 
the reflected Bogoliubov waves is scanned.
The linear potential which has been used frequently near a turning point
has the peculiar feature that within our numerical accuracy, 
its phase shift vanishes.
In other potentials like a hard wall or an exponential soft barrier, 
the reflection is dispersive, i.e., the phase shift depends on energy.
%• • • 
%Since a parabolic potential is so common, consider also a “quadratic
%quarter pipe” with finite slope. Corresponds to fixed Thomas-Fermi radius,
%but keep potential at zero beyond it (avoid mode quantisation). 
%• • •
These results cannot be understood 
%within semiclassical mechanics
%(geometrical optics) 
by considering only the potential barrier 
because of the mean-field potential due to the condensate density.
Our analysis reveals in particular certain contributions to the
phase shift due to the spatially inhomogeneous condensate density, 
i.e., contributions beyond the local density approximation.

The basic equations of Bogoliubov theory are recalled in Sec.\:\ref{s:model}
where also exact reference solutions are given with respect to which the
phase shift may be defined. 
Sec.\:\ref{s:analysis} outlines the main analysis of the problem and
recalls an adiabatic scheme that eliminates divergence problems approximately.
Nonadiabatic corrections are dealt with in Subsec.\:\ref{s:phase-shift-2},
the particular case of the linear potential in Subsec.\:\ref{s:linear-phase}.
Applications are discussed in Sec.\:\ref{s:applications},
while more technical material and further plots have been relegated 
to the Appendices.
In a companion paper \citep{phase-shift-II}, the Bogoliubov modes are
used to compute quantum and thermal corrections to the density profile.
A brief analysis of the zero-point energy density of the Bose gas has been 
given in \citet{Henkel25b_arxiv}.
We find enhanced fluctuations in the 
border region around the turning point, whose dependence on temperature
is quite different compared to a homogeneous system. 
This illustrates the fallacies of the common implicit assumption 
behind the local density approximation.
The issue of quantum fluctuations in the atomic density is of interest
to study non-trivial quantum vacuum effects in laboratory settings,
see, e.g., \citet{Roberts05, Edmonds23}.
The turning point problem is also similar to the electron density 
near the surface of a conductor \citep{Lang69, Kenner72} 
and its excitations, for example
in the form of surface plasmons \citep{Feibelman82, Apell84b, Schaich85,
Liebsch_Book}.
This problem has recently revived some interest in hydrodynamic models
for metallic electrons \citep{Forcella14, Toscano15, Yan15b, Benedicto15, Ciraci16, Yang19c, Mortensen21}.

\begin{figure}[tbh]
\centerline{%
\includegraphics*[height=0.3\textwidth]{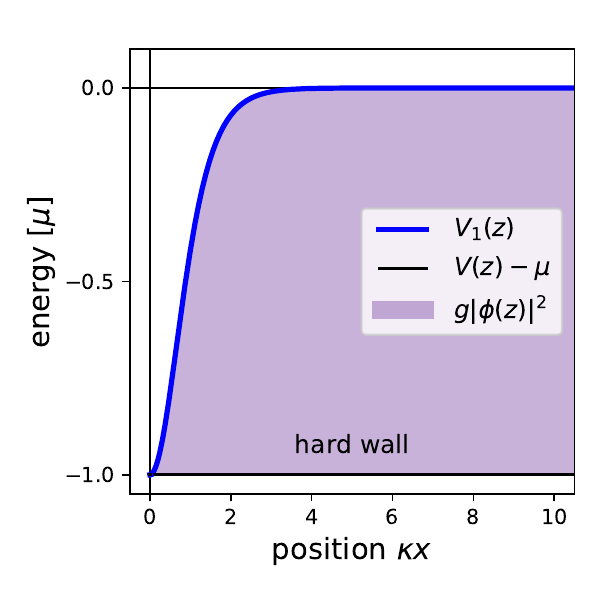}%
\hspace*{-1.5mm}%
\includegraphics*[height=0.3\textwidth]{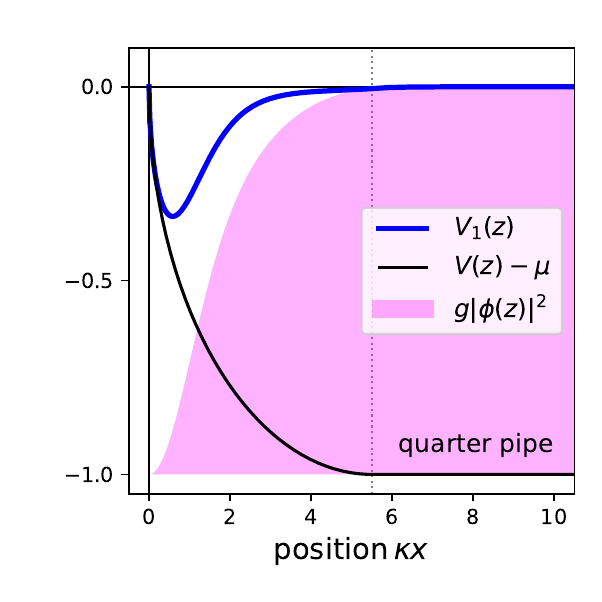}%
% measured on screen: hard 8,15 vs half 8,85, 0,3*8,15/8,85 =
% \raisebox{1.5mm}{\includegraphics*[height=0.278\textwidth]{../Figs/sketch_HalfPipe_2.pdf}}%
}

\centerline{%
\includegraphics*[height=0.3\textwidth]{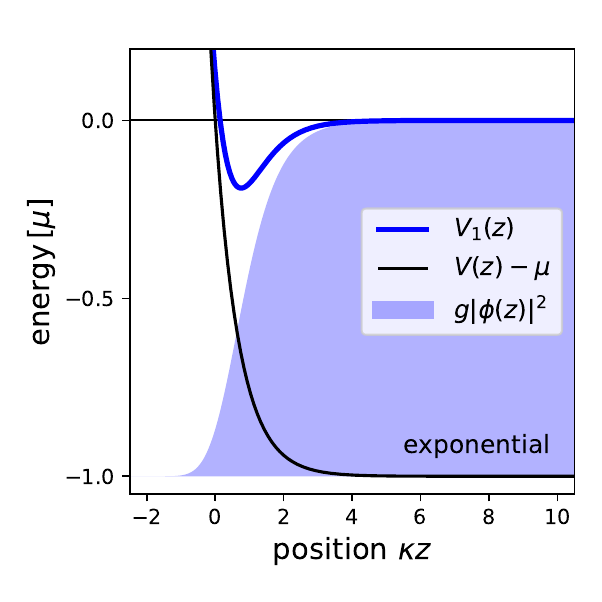}%
\hspace*{-1.5mm}%
\includegraphics*[height=0.3\textwidth]{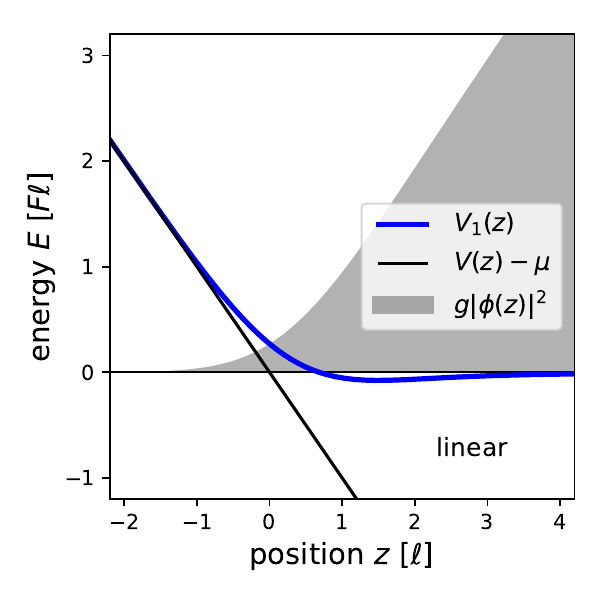}%
% measured on screen: exp: 7.55, lin: 8.05, 0.3*7.55/8.05 = 
%\raisebox{1.0mm}{\includegraphics*[height=0.2814\textwidth]{../Figs/sketch_Linear_2.pdf}}%
}

\caption[]{%
%• • • 
%add a “half harmonic” potential, left open after $V = 0$.
%• • •
Kaleidoscope of potentials and condensate densities considered 
in this paper. From top left to bottom right: hard-wall potential $V = 0$,
quarter-pipe potential that connects smoothly a hard wall at $z = 0$ to a
flat bottom at $z \ge R$ (dotted line),
exponential $V(z) = \mu \exp(- a z)$ and linear $V(z) = \mu - F z$
potential. 
%The right column provides the response of the condensate (and its density)
%to a change in the chemical potential: the function 
%$\chi = \partial \phi / \partial \mu$ and 
%$\partial n / \partial \mu = 2 \phi \chi$.
The function $V_1 = V - \mu + g \phi^2$ (thick blue lines)
applies to the elementary excitations of the condensate phase, 
see Eq.\,(\ref{eq:BdG-phase}).
\\
Energies are scaled to the chemical potential $\mu$, except for 
the linear potential where, since $\mu = 0$ by a suitable choice of coordinates,
the natural unit $F \ell$ [see Table\,\ref{t:units}] is used.
The turning point is at $z = 0$ in all cases, although the boundary conditions
there depend on the type of potential (Dirichlet $\phi(0) = 0$ for the hard-wall
potentials in the top row).
Position coordinate scaled by the inverse healing length $\kappa$,
as defined in Eq.\,(\ref{eq:tanh-condensate}) and Table~\ref{t:units}.
%\\
%Extend this with “natural symmetry transformations” and their mode functions:
%phase fluctuations = $\phi$ itself, ground state of $H_1$; 
%shift in chemical potential = $\chi$,
%solution of inhomogeneous equation with $H_3$, corresponding density profile
%(surprisingly similar because flat away from turning point).
}
\label{fig:sketches}
\end{figure}

\section{Model}
\label{s:model}

We consider a quasi-one-dimensional Bose condensate and focus on its border,
i.e., the region where the density smoothly drops to zero. 
This can be implemented with a hard or a soft barrier: 
the former case is equivalent to a Dirichlet boundary condition. 
The collective wave function then rises from zero to a constant density, 
on the scale of the healing length 
defined by the nonlinearity of the wave equation, 
i.e., particle interactions at the mean-field level.
A soft barrier, on the other hand, introduces the same length scale 
as the one-particle Schrödinger equation,
where, e.g., the Airy function provides a smooth connection
to the tunnelling regime below the barrier. 
In the physical solution to the nonlinear wave equation 
for the condensate [Eq.\,(\ref{eq:GPe}) below],
the nonlinearity becomes dominant on the dense side of the turning point 
where it suppresses the oscillatory behaviour of the Airy function.

\subsection{Overview}

The potential barriers we discuss in this paper are illustrated in 
Fig.\,\ref{fig:sketches}. 
Top left is shown a flat potential with a hard wall 
at $z = 0$. This case provides an analytic solution given by a “dark
soliton” \citep{PitaevskiiStringari, Clark97, Konotop08}. 
On the top right, we show an intermediate case
where the hard wall and a flat bottom are smoothly connected by an
elliptical arc. We call this potential the “quarter pipe”. 
Bottom left is a smooth barrier that decays exponentially
to a flat bottom. 
Finally, a linear potential is shown bottom right: here,
the Thomas-Fermi condensate density grows linearly with distance from the
turning point (shaded area). 
The linear potential is the starting point for Langer's analysis of the
WKB connection problem in wave mechanics \citep{MessiahI}. 
The nonlinear wave equation then connects to a celebrated mathematical problem
with movable singularities, the Painlevé~II equation \citep{DLMF, Dalfovo96}.

We choose coordinates such that $z = 0$ corresponds either to the position 
of the Dirichlet boundary condition (hard wall and quarter pipe) or to the
crossing of the trapping potential $V(z)$ and the chemical potential $\mu$
(exponential and linear cases).
The chemical potential $\mu$ provides the “healing length” $1/\kappa$ 
as a natural length scale, see Table~\ref{t:units}. 
Typical parameters for all cases are listed in Table~\ref{t:units}.
The linear potential is the local version of a soft wall, but also applies
in a homogeneous gravitational field. 
For a natural length unit in this case, combine the force $F = -dV/dz$ 
and the particle mass $M$ into $\ell = (\hbar^2 / 2 M F)^{1/3}$.
Interactions between particles in ultracold atomic gases 
can be parametrised by a single parameter $g$ 
proportional to the s-wave scattering length \citep{DrakeHandbook}.
In quasi-one-dimensional trapping geometries, it also depends 
on the transverse confinement \citep{Olshanii98, BouchouleChapter11};
generic values are $g \sim 0.3 \ldots 3\,{\rm nK}\,\upmu{\rm m}$.
Typical values for the length scales $1/\kappa$ and $\ell$ 
are in the sub-micron range, see also Table~1 in \citet{Diallo15a}.

\begin{table}[t!bh]
\begin{center}
\begin{tabular}{l|llll}
& hard wall & quarter pipe & exponential & linear
\\
\hline
potential $V(z)$ &
$0$ & $\mu - \mu \sqrt{(2 - \frac zR) \frac zR}$ & 
$\mu\,\exp(- a z)$ & $\mu - F z$
% quarter-pipe
% mu - (mu/R)sqrt( R^2 - (x-R)^2 )
% = mu - mu sqrt(1 - (1 - x/R)^2)
% = mu - mu sqrt((2 - x/R) x/R)
\\
length scale(s) &
$\frac{1}{\kappa} = (\frac{\hbar^2}{\mu M})^{1/2}$ &
range $R$ &
decay length $1/a$ &
$\ell = (\frac{\hbar^2}{2 M F})^{1/3}$
\\
value & $230\,{\rm nm}$ 
& $R = 1 \ldots 1.8\,\mbox{µ}{\rm m}$
& $1/a = 50\ldots 200\,{\rm nm}$ 
& $\ell = 300\,{\rm nm}$
\\
parameters & $\mu = 100\,{\rm nK}$
& 
& 
& $F/M = 9.81\,{\rm m/s}^2$
\\
numerics
& $\mu = 1$
& $\kappa R = 2 \ldots 10$
% & $a = 0.3 \ldots 3/\mu$
% OK, this is how we started to define a (having the slope of the
% potential in mind). But in the rest of the paper, we always compare
% a and kappa, and for mu = 1, kappa = 1/sqrt(2). So we write
& $\kappa/a = 0.24 \ldots 2.4$
& $\mu = 0$, $F = 1 = \ell$
\end{tabular}
\end{center}
\caption[]{Characteristic units and parameters. The numbers (atom mass $M$,
resonant wavelength) are evaluated for the Rb-87 atom.
The quarter-pipe potential continues as $V(z) = 0$ for $z \ge R$.
The exponential potential can be implemented with a blue-detuned light field
created by total internal reflection (evanescent wave). 
Its length scale $1/a$ is comparable to the Rb D2 wavelength $780\,{\rm nm}$, 
but depends on optical parameters.
% For the numerics, we use units with $\hbar^2/2M = 1$.
}
\label{t:units}
\end{table}

\subsection{Basic equations}

Adopt suitable units, we set $\hbar^2 / 2 M = 1$, 
and the nonlinear Schr\"odinger equation for the condensate $\phi$, 
also known as Gross--Pitaevskii equation \citep{PitaevskiiStringari},
takes the form 
%known as the second
%Painlev\'e transcendent~\citep{Ablowitz77, Dalfovo96, Lundh97, Fetter98a}:
\begin{equation}
- \frac{ {\rm d}^2 \phi }{ {\rm d}z^2 } + V(z)\, \phi + g |\phi|^2 \phi = \mu \phi
	\label{eq:GPe}
\end{equation}
where the constant $g$ scales the mean-field potential $g |\phi|^2 > 0$.

In Fig.\,\ref{fig:sketches}, 
numerical results for the condensate density (obtained as outlined in
Sec.\:\ref{s:numerics} below) are shown.
They illustrate a smooth cross-over
between vanishingly small values for $z \leq 0$ and the Thomas-Fermi profile
\begin{equation}
z \to \infty: \quad g |\phi_{\rm TF}|^2(z) = \mu - V(z)
\label{eq:Thomas-Fermi-density}
\end{equation}
The latter is obtained by neglecting the derivative in Eq.\,(\ref{eq:GPe}).
Typical atom numbers within the healing length $1/\kappa$
are of the order of $\mu / (g \kappa) \sim 5 \ldots 50$.
The mean-field theory behind the Gross-Pitaevskii framework is valid if
this number is not small compared to unity \citep{Mora03, Kerr24}.
A more quantitative comparison in terms of “missing particles” in the
region 
around the turning point $z = 0$ is provided in the companion paper
\citep{phase-shift-II}, Appendix A.1. % \ref{a:dNc_border}.
%[see \citet{phase-shift-II},
%Eq.\,(\ref{eq:dNc-for-dark-soliton}) in \ref{a:dNc_border}].

The linearization around the mean field,
$\phi \mapsto \phi + u\, {\rm e}^{ - {\rm i} E t } + v^*\, {\rm e}^{ {\rm i} E t }$
leads to the (stationary) Bogoliubov--de Gennes equations 
that read, in the same units,
\begin{eqnarray}
{-} \frac{ {\rm d}^2 u }{ {\rm d}z^2 } + V(z) \, u + 2 g |\phi|^2 \, u
+ g \, \phi^2 \, v &=& (\mu + E) \, u
\,,
\nonumber\\
{-} \frac{ {\rm d}^2 v }{ {\rm d}z^2 } + V(z) \, v + 2 g |\phi|^2 \, v
+ g \, \phi^{* 2} \, u &=& (\mu - E) \, v
\,.
	\label{eq:BdG}
\end{eqnarray}
We fix the phases of $\phi, u, v$ to be real, choosing positive $\phi$. 
We also restrict the spectrum to $E > 0$, since to a pair $(u, v)$ with
energy $E$ corresponds another pair $(v, u)$ with $-E$.

The sum and difference of the Bogoliubov amplitudes physically
represent density and phase quadratures of the condensate, 
as can be seen from
$u\, {\rm e}^{ - {\rm i} E t } + v^*\, {\rm e}^{ {\rm i} E t }
= (u + v) \cos E t - {\rm i} (u - v) \sin E t$. 
Defining the family of effective one-particle Hamiltonians
\begin{equation}
H_n = - \frac{ {\rm d}^2 }{ {\rm d}z^2 } + V(z) - \mu + n \, g \phi^2
\,,
\qquad (n = 1, 2, 3)
\,,
	\label{eq:def-H123}
\end{equation}
we get from Eqs.\,(\ref{eq:BdG}) the equation pair
\begin{eqnarray}
H_1 \phiMinus &=& E \fPlus
\,,
\qquad 
\phiMinus = { (u - v) }{ /\sqrt{2} }
\,,
\nonumber
\\
H_3 \fPlus &=& E \phiMinus
\,,
\qquad 
\fPlus = { (u + v) }{ /\sqrt{2} }
\,.
	\label{eq:BdG-phase}
\end{eqnarray}
The potential $V_1 = V - \mu + g \phi^2$ is shown in thick blue in 
Fig.\,\ref{fig:sketches}.
It quantifies the deviation of the condensate
density from its Thomas-Fermi approximation and therefore depends
on its spatial variation;
at large distance from the turning point, deep in the condensate,
$V_1(z) \to 0$, 
while $V_1(z) \to V - \mu$ on the dilute side.
% A third representation based on local (or adiabatic) eigenvectors of 
% the $2\times 2$ system has been introduced by \citet{Diallo15a},
% see Sec.\,\ref{s:adiabatic} below.

\subsection{Exact solutions}

For a hard-wall potential, $V(z) = 0$ for $z > 0$,
and a Dirichlet boundary condition $\phi(0) = 0$ is imposed 
(also on $u$ and $v$). 
In this simple case, exact solutions are known:
the so-called “dark soliton” provides a condensate wave function that
goes through zero \citep{PitaevskiiStringari}.
Cutting this solution in half, the Dirichlet boundary condition case 
is solved by
\begin{equation}
\phi(z) = \sqrt{\mu/g} \, \tanh( \kappa z )
\,,\qquad
% \kappa = \frac{\sqrt{2 M g n} }{\sqrt{2} \hbar }
\kappa = \sqrt{\mu/2} 
= \sqrt{M \mu} / \hbar
\,.
\label{eq:tanh-condensate}
\end{equation}
The Bogoliubov mode functions can also be found explicitly
and read \citep{Kovrizhin01a, Dziarmaga04, Negretti08a, Walczak11}
\begin{eqnarray}
u_k(z) =& 
%{\cal N_k} \, {\rm e}^{ {\rm i} k z } \left\{
%\frac{ k }{ \cosh^2( \kappa z ) }
%+ \alpha_k \left[ 
%k + 2 {\rm i} \kappa \tanh( \kappa z ) 
%\right]
%\right\}
%{\cal N_k} \left\{ 
A_k \, k \sin(k z)
\left[ 
\frac{ 1 }{ \cosh^2( \kappa z ) }
+ \alpha_k
\right]
+ 2 A_k \, \alpha_k 
\kappa \cos(k z) \tanh( \kappa z ) 
%\right\}
\nonumber
\\
v_k(z) =& 
%{\cal N_k} \, {\rm e}^{ {\rm i} k z } \left\{
%\frac{ k }{ \cosh^2( \kappa z ) }
%+ \beta_k \left[ 
%k + 2 {\rm i} \kappa \tanh( \kappa z ) 
%\right]
%\right\}
%{\cal N_k} \left\{ 
A_k \, k \sin(k z)
\left[ 
\frac{ 1 }{ \cosh^2( \kappa z ) }
- \beta_k
\right]
- 2 A_k \, \beta_k 
\kappa \cos(k z) \tanh( \kappa z ) 
%\right\}
\label{eq:uv-hard-wall}
\\
\left.
\begin{array}{c}
\alpha_k \\ \beta_k
\end{array}
\right\} & = % \frac{ k^2 }{ \kappa^2 } \pm \frac{ 2 E }{ \mu }
%\frac{ k^2 \pm E }{ 2 \kappa^2 }
\frac{ E \pm k^2 }{ \mu }
\label{eq:def-alpha-beta}
\end{eqnarray}
The energy is given by the Bogoliubov dispersion relation
\begin{equation}
% E & = \hbar c |k| \sqrt{ 1 + (k/2\kappa)^2 }
% speed of sound m c^2 = g n = \mu, hence in our units
% c = sqrt{2 \mu}
%E = |k| \sqrt{ 4 \kappa^2 + k^2 }
E = |k| \sqrt{ 2 \mu + k^2 }
\label{eq:Bogoliubov-dispersion}
\end{equation}
We choose the normalisation $A_k$
%${\cal N}$ 
% find explicit form for A_k: asymptotic form of 
% u(z) =   A_k alpha_k [k sin kz + 2 kappa cos kz]
% v(z) = - A_k beta_k  [k sin kz + 2 kappa cos kz]
% so common phase shift is tan delta = 2 kappa / k
% squared prefactors to remove theta:
% A_k^2 (alpha_k^2 + beta_k^2) = 1/k
% and alpha_k^2 + beta_k^2 = 2 (E^2 + k^4) / mu^2 
%     = 4 k^2 (mu + k^2) / mu^2
% A_k = mu / [ 2 sqrt(k^3 (mu + k^2)) ]
%
such that in the region $\kappa z \gg 1$ 
where the condensate $\phi(z)$ is constant [see Eq.\,(\ref{eq:tanh-condensate})], 
we have
%• • • 
%check carefully: perhaps the $\delta$ convention as in standard scattering
%phases is more natural. Write $j = \sin$ and $y = -\cos$ (opposite sign),
%check standard DLMF definition of Wronskian
%$W[j, y] = j y' - j' y = +1$ (conforms with Bessel-Coulomb)
%and extract from $u = j \cos(\delta) - y \sin(\delta)$ with
%$\cos(\delta) = W[u, y]$ and
%$\sin(\delta) = W[u, j]$
%This gives (only difference so far!) $u = \sin(kz + \delta)$
%• • •
\begin{equation}
u_k(z) \to \frac{ \cos(\theta_k/2) }{ \sqrt{k} } \sin(k z + \delta)
\,,\qquad
v_k(z) \to -\frac{ \sin(\theta_k/2) }{ \sqrt{k} } \sin(k z + \delta)
\label{eq:asymptotic-u-and-v-0}
\end{equation}
The relative amplitudes of $u_k$ and $v_k$ are set by the angle 
$\theta_k$
that we read off from Eqs.\,(\ref{eq:uv-hard-wall}, \ref{eq:def-alpha-beta}):
\begin{equation}
\tan\theta_k = \frac{ 2 \alpha_k \beta_k }{ \alpha_k^2 - \beta_k^2 }
= \frac{ \mu }{ E } 
\label{eq:theta-angle-hw}
\end{equation}
With this choice, the wave functions are orthogonal in the
continuous spectrum, as has been shown in \citet{Diallo15a}
\begin{equation}
\int\limits_{0}^{\infty}\!\frac{ {\rm d}z }{ \pi } 
\left[ u_k(z) u_{k'}(z) - v_k(z) v_{k'}(z) \right]
= \delta( E - E' )
\label{eq:uv-continuum-normalisation}
\end{equation}
And finally, the phase shift $\delta$ % in Eq.\,(\ref{eq:asymptotic-u-and-v-0})
is given by
\begin{equation}
% cos delta = k, sin delta = 2 kappa 
\tan \delta = \frac{2\kappa}{k}
\label{eq:delta-hw}
\end{equation}
This phase shift will be the main focus of this paper. 
As can be seen in Eq.\,(\ref{eq:asymptotic-u-and-v-0}),
it describes
the position of the nodes of the $u$ and $v$ wave functions,
relative to the reference position $z = 0$ set by the hard-wall boundary condition.
For an illustration, see Fig.\:\ref{fig:uv_adiabatic} below.

Another exact solution is available in the case of the linear potential.
The Gross-Pitaevskii Eq.\,(\ref{eq:GPe})
is solved by the Painlevé~II transcendent \citep{Ablowitz77, Hastings80, Dalfovo96}.
It interpolates at low densities to
the Airy function familiar from the Schrödinger equation in the linear potential,
\begin{equation}
z \to -\infty: \quad
\phi^2(z) \to 2\, {\mathop{\rm Ai}}^2(-z)
\label{eq:Airy-asymptote}
\end{equation}
(Here and in the following, we use the natural units $F = 1 = \ell = g$, 
see Table~\ref{t:units}.)
The scale factor $2$ is not arbitrary since we are dealing with a
nonlinear wave equation; 
it is characteristic for that Painlevé~II solution 
that connects on the other side of the turning point to the Thomas-Fermi 
approximation~(\ref{eq:Thomas-Fermi-density}).
It has been shown by \cite{Hastings80} that if the prefactor in
Eq.\,(\ref{eq:Airy-asymptote}) is larger than $2$, then the solution
diverges for some finite value of $z > 0$. If it is smaller, then
it has an oscillatory asymptote
\begin{equation}
z \to \infty: \quad \phi^2(z) \approx
\frac{ 2 d^2 }{ z^{1/2} } \sin^2\left( 
\tfrac23 z^{3/2} - \tfrac34 d^2 \log z - c 
\right)
\label{eq:}
\end{equation}
with some constants $d, c$. 
If the prefactor, say $a^2$, in Eq.\,(\ref{eq:Airy-asymptote}) approaches
$2$, \citet{Ablowitz77} show that the constant $d$ scales as
\begin{equation}
d^2 \approx \frac{ -1}{ \pi } \log(1 - a^2/2)
\label{eq:}
\end{equation}
Coming back to the Thomas-Fermi behaviour of Eq.\,(\ref{eq:Thomas-Fermi-density}),
expanding $\phi(z)$ around it and inserting this into the Gross-Pitaevskii
equation, finally yields the post-Thomas-Fermi series 
\citep{Lundh97, Margetis00}
\begin{equation}
z \to \infty: \quad \phi^2(z) =
z - \frac{ 1 }{ 4 z^2 } - \frac{ 9 }{ 8 z^5 } + {\cal O}(z^{-8})
\label{eq:post-TF-condensate-density}
\end{equation}
With the help of this expansion, we find that
the effective potential $U_{\rm ad}(z)$ introduced in 
Eq.\,(\ref{eq:BdG-adiabatic-phi}) below approaches the asymptote
\begin{equation}
z \to \infty: \quad
U_{\rm ad}(z) \to 
- \frac{E^2 }{2 z}
- \frac{1}{4 z^2} 
\,.
	\label{eq:long-range-Coulomb-potential}
\end{equation}
% where the two terms arise from $D(z)$ and $V_1(z)$, respectively.
These slowly decaying terms constitute a reference potential 
that is familiar from the Coulomb interaction
and centrifugal barriers, although the latter features a strange sign.
It has exact solutions right at the continuum threshold given by 
ordinary Bessel functions \citep{DLMF}
%[• sign to check: 
%our Wronskian (30) is now following the DLMF convention. 
%Their formula 10.5.2 gives
%$W[J_0(x), Y_0(x)] = 2 / (\pi x)$ and therefore
%$W[j, y] = (\pi z) (E / \sqrt{2z}) (2/(\pi E \sqrt{2z}) = 1$.
%This is why we redefined the plane wave/flat bottom solutions 
%to $y = - \cos$ so that we get exactly the same convention •]
\begin{equation}
j_E(z) = \sqrt{\pi z} \, J_0( E \sqrt{2z} )
\,, \qquad
y_E(z) = \sqrt{\pi z} \, Y_0( E \sqrt{2z} )
\label{eq:Bessel-Coulomb-solutions-0}
\end{equation}
The first solution stays regular when extrapolated to $z \to 0$, while
the second one has a divergent slope there.
We have chosen a normalisation such that the Wronskian determinant
of the two solutions is $W[j_E, y_E] = 1$. 
The Wronskian will turn out a useful tool to project a general solution 
onto a pair of basis functions. 
It is defined as \citep{DLMF, BenderOrszag}
\begin{equation}
W[u, w] 
= u \frac{ {\rm d} w }{ {\rm d}z } - \frac{ {\rm d} u }{ {\rm d}z } w 
\,.
\label{eq:def-Wronskian}
\end{equation}
The actual solution $\uTilde(z)$ in the potential $U_{\rm ad}(z)$
is, for large $z$, a linear combination of reference waves. It gives
access to the asymptotic behaviour of the Bogoliubov amplitudes $u(z)$, 
$v(z)$ [see Eq.\,(\ref{eq:rotated-basis}) below]
and thus defines the phase shift $\delta$
%[•
%formula stays right
%•]
\begin{equation}
z \to \infty: \qquad
\uTilde(z) \to 
j_E(z) \cos\delta - y_E(z) \sin\delta
\label{eq:def-phase-2}
\end{equation}
More explicitly, a Wronskian yields $W[\uTilde, j_E] \to A \sin\delta$.
% The continuum normalisation we adopt is $A = 1$.
According to the large-argument expansion of the Bessel functions \citep{DLMF},
we get, similar to Eq.\,(\ref{eq:asymptotic-u-and-v-0}), 
%$x \gg 1/(2 E^2)$
\begin{eqnarray}
%j(x) &\to& \frac{ (2 x)^{1/4} }{ \sqrt{ E } } \cos( E \sqrt{2x} - 3\pi/4)
%\\
%y(x) &\to& \frac{ (2 x)^{1/4} }{ \sqrt{ E } } \sin( E \sqrt{2x} - 3\pi/4)
%J(x) &\to& \frac{ (2 x)^{1/4} }{ \sqrt{ E } } \cos( E \sqrt{2x} - \pi/4)
%\\
%Y(x) &\to& \frac{ (2 x)^{1/4} }{ \sqrt{ E } } \sin( E \sqrt{2x} - \pi/4)
%\\
E \sqrt{ 2 z } \gg 1: \qquad
\uTilde(z) & \approx & 
% sign change when going from cos to sin checked
% A 
\frac{ (2 z)^{1/4} }{ \sqrt{ E } } \sin( E \sqrt{2 z} + \pi/4 
	+ \delta)
\,.
\label{eq:asymptotic-Bessel-Coulomb}
\end{eqnarray}
The square root in the sine argument makes the wavelength increase at large $z$,
in distinction to the constant wavelength of Eq.\,(\ref{eq:asymptotic-u-and-v-0})
that applies to “flat-bottom” potentials.

\subsection{Numerics}
\label{s:numerics}

For numerical solutions of the Gross-Pitaevskii and Bogoliubov equations,
see, e.g., \cite{Xie09b, Gao20, Sadaka24}. 
These approaches also cover two- and three-dimensional systems and are
mostly based on finite-element techniques. 
In the one-dimensional system considered here, 
the condensate solution can be found by minimising the Gross--Pitaevskii
energy functional (units with $\hbar^2 / 2m = 1$)
\begin{equation}
{\cal F}[\phi] = \int\!{\rm d}z\left\{
\frac{1}{2} \Big( \frac{ {\rm d}\phi }{ {\rm d}z } \Big)^2
+ \frac{1}{2} (V - \mu) \phi^2 + \frac{g}{4} \phi^4 
\right\}
\label{eq:GPe-functional}
\end{equation}
When represented on a discrete grid, the \texttt{minimize} 
routine of scipy may be used
\citep{scipy_minimize}.
To improve the solution, we iterate the following recursive scheme
\begin{equation}
\delta \phi^{(n)} = H_1^{(n)} \phi^{(n)} \,,
\quad
H_3^{(n)} \chi^{(n+1)} = - \delta \phi^{(n)} \,,
\quad
\phi^{(n+1)} = \phi^{(n)} + \chi^{(n+1)}
\label{eq:recursive-H1-H3-scheme}
\end{equation}
where the mean-field potentials in $H_1^{(n)}, H_3^{(n)}$ are evaluated 
based on $|\phi^{(n)}|^2$.
Convergence is achieved when the norm of the error $\delta \phi^{(n)}$ 
in the computational domain drops below a predefined precision.

The Bogoliubov equations~(\ref{eq:BdG})
and the iterative problem~(\ref{eq:recursive-H1-H3-scheme})
may be solved with sparse linear algebra packages \citep{scipy_minimize},
using a finite-difference scheme 
and suitable boundary conditions at both ends: 
a Dirichlet boundary condition 
at the left end (taken at $z = 0$ or sufficiently deep in the tunnelling 
domain), 
and Dirichlet or Neumann conditions on the right. 
This yields a set of discrete eigenvalues 
that is getting denser when the computational domain is enlarged.
To access the continuous spectrum with a smaller domain and 
in particular at very low energies,
we use a Robin boundary condition, i.e., 
enforcing the logarithmic derivative 
$Z = (1/u) \, {\rm d}u / {\rm d}z$ at the last grid point $z = L$.
This can be done by adjusting the second derivative $u_{zz}$
on the grid. 
With a grid spacing ${\rm d}z$, we take at $z = L$
\begin{eqnarray}
u_{zz}(L) &= \frac{ 1 }{ {\rm d}z } 
\Big[ 
%Z \, u(L) - \frac{ u(L) - u(L - {\rm d}z) }{ {\rm d}z } 
\frac{ u(L - {\rm d}z) - u(L) }{ {\rm d}z }
+ Z \, u(L) 
\Big]
\nonumber\\
& =
\frac{ 1 }{ {\rm d}z^2 } \big[
u(L - {\rm d}z) - ( 1 - Z \, {\rm d}z ) \, u(L) 
\big]
\label{eq:}
\end{eqnarray}
When a Dirichlet (or Neumann) boundary condition is applied
at $z = L$,
the number $-(1 - Z \, {\rm d}z)$ is replaced by $-2$ (or $-1$).
(In the interior of the grid, the three-point stencil 
$(1, -2, 1)/{\rm d}z^2$ is used.)
The best accuracy is obtained when $Z$ is
calculated with a derivative 
${\rm d}u / {\rm d}z$ evaluated at mid-point, 
i.e., at $L + {\rm d}z/2$ just outside the computational grid.
The logarithmic derivative can also implement boundary 
conditions for tunnelling solutions,
computed according to the WKB approximation \citep{MessiahI}.
We have checked that the scheme is compatible with compact fourth-order 
finite-difference schemes \citep{Lele92, Xie09b} that improve the
accuracy in ${\rm d}z$ by two orders.

\section{Analysis}
\label{s:analysis}

\subsection{Adiabatic basis}
\label{s:adiabatic}

The eigenproblem formulation sketched in the previous section is facing 
the issue that the Bogoliubov equations~(\ref{eq:BdG})
contain at least one unstable mode that grows exponentially fast. 
To make analytical and numerical progress,
an adiabatic basis has been introduced by \citet{Diallo15a}. 
It is given by the eigenvectors of the $2\times 2$ matrix $\mathbf{M}$
underlying the Bogoliubov system
[see Eq.\,(\ref{eq:def-Bogoliubov-matrix-M}) below].
For an intuitive understanding of the relevant transformation, 
we introduce the canonical “position” and 
“momentum” variables $(u, v, p = {\rm d}u/{\rm d}z, q = {\rm d}v/{\rm d}z)$
and consider a joint rotation in the $uv$- and $pq$-planes:
%\begin{equation}
%\left(\begin{array}{c}
%\uTilde\\
%\vTilde\\
%\tilde{p}\\
%\tilde{q}
%\end{array}\right)
%=
%\left(\begin{array}{cccc}
%\cos(\theta/2) & -\sin(\theta/2) & & \\
%\sin(\theta/2) & \cos(\theta/2) & & \\
%& & \cos(\theta/2) & -\sin(\theta/2) \\
%& & \sin(\theta/2) & \cos(\theta/2)
%\end{array}\right)
%\left(\begin{array}{c}
%{u}\\
%{v}\\
%{p}\\
%{q}
%\end{array}\right)
%\end{equation}
\begin{eqnarray}
\left(\begin{array}{c}
\uTilde\\
\vTilde\\
\end{array}\right)
&=&
\mathbf{R}
\left(\begin{array}{c}
{u}\\
{v}\\
\end{array}\right)
=
\left(\begin{array}{cccc}
\cos(\theta/2) & -\sin(\theta/2) \\
\sin(\theta/2) &  \cos(\theta/2)
\end{array}\right)
\left(\begin{array}{c}
{u}\\
{v}\\
\end{array}\right)
\nonumber
\\
\left(\begin{array}{c}
\tilde{p}\\
\tilde{q}
\end{array}\right)
&=&
\mathbf{R}
\left(\begin{array}{c}
{p}\\
{q}
\end{array}\right)
\label{eq:rotated-basis}
\end{eqnarray}
In the special case $\theta = \pi/2$, 
one recovers the difference and sum modes $\phiMinus$,
$\fPlus$ of Eq.\,(\ref{eq:BdG-phase}).
Our notation conveys that $\uTilde$ shares many similarities
with the phase mode $\varphi_-$, while $\vTilde$ is similar 
to the the density mode $f_+$. 
The rotation into the adiabatic basis requires, however,
a position-dependent angle $\theta(z)$. 

Keeping $\theta$ dependent on position,
computing the derivatives and using the Bogoliubov equations~(\ref{eq:BdG}), 
we find the following set of equations in this representation
% (• • • sign of $\theta'$ OK • • •)
%
%\begin{equation}
%\frac{ {\rm d} }{ {\rm d}z } 
%\left(\begin{array}{c}
%\uTilde\\
%\vTilde\\
%\tilde{p}\\
%\tilde{q}
%\end{array}\right)
%=
%\left(\begin{array}{cc}
%\begin{array}{cc}
%& \theta'/2 \\
%- \theta'/2 &
%\end{array}
%&
%\begin{array}{cc}
%1 & \\
%& 1
%\end{array}
%\\
%R^{-1} B R 
%& 
%\begin{array}{cc}
%& \theta'/2 \\
%- \theta'/2 &
%\end{array}
%\end{array}\right)
%\left(\begin{array}{c}
%\uTilde\\
%\vTilde\\
%\tilde{p}\\
%\tilde{q}
%\end{array}\right)
%\end{equation}
%
%
%
\begin{eqnarray}
\frac{ {\rm d} }{ {\rm d}z } 
\left(\begin{array}{c}
\uTilde\\
\vTilde
\end{array}\right)
& = &
\left(\begin{array}{c}
\tilde{p}\\
\tilde{q}
\end{array}\right)
+ 
\left(\begin{array}{cc}
0 & - \theta'/2 \\
\theta'/2 & 0
\end{array}\right)
\left(\begin{array}{c}
\uTilde\\
\vTilde
\end{array}\right)
\nonumber
\\
\frac{ {\rm d} }{ {\rm d}z } 
\left(\begin{array}{c}
\tilde{p}\\
\tilde{q}
\end{array}\right)
& = &
\mathbf{R} \mathbf{M} \mathbf{R}^{-1}
\left(\begin{array}{c}
\uTilde\\
\vTilde
\end{array}\right)
+
\left(
\begin{array}{cc}
0 & - \theta'/2 \\
\theta'/2 & 0
\end{array}\right)
\left(\begin{array}{c}
\tilde{p}\\
\tilde{q}
\end{array}\right)
\label{eq:BdG-adiabatic-basis-0}
\end{eqnarray}
The derivatives $\theta' = {\rm d}\theta / {\rm d}z$ 
appear % in Eqs.\,(\ref{eq:BdG-adiabatic-basis-0})
as off-diagonal elements.
This cross-coupling between “positions” and “momenta” 
is reminiscent of the distinction between kinematic and canonical momenta
for a charged particle in two dimensions 
in a magnetic field $B \propto \theta'$.
In the second line of Eq.\,(\ref{eq:BdG-adiabatic-basis-0}), 
$\mathbf{M}$ is the Bogoliubov matrix
\begin{equation}
\mathbf{M} = \left( \begin{array}{ll}
V_2(z) - E & g \phi^{2}(z) 
\\
g \phi^{2}(z) & V_2(z) + E
\end{array} \right)
\label{eq:def-Bogoliubov-matrix-M}
\end{equation}
with $V_2(z) = V(z) - \mu + 2 g |\phi(z)|^2$.
It becomes diagonal when the angle $\theta$ in the rotation matrix $\mathbf{R}$ 
is chosen as
\begin{equation}
\tan\theta(z) = \frac{g\phi^2(z)}{E}
\label{eq:def-adiabatic-angle}
\end{equation}
This is the position-dependent generalisation of 
Eq.\,(\ref{eq:theta-angle-hw}). 
The eigenvalues of $\mathbf{M}$ are
\begin{eqnarray}
%R M R^{-1} &= 
%\left( \begin{array}{cc}
%U_{\rm ad} & \\
%& V_{\rm ad}
%\end{array} \right)
%\\
\left. \begin{array}{l}
%U_{\rm ad}(z) \\
%V_{\rm ad}(z)
M_{\uTilde}(z) \\
M_{\vTilde}(z)
\end{array} \right\} & = &
V_2(z) \mp \sqrt{E^2 + g^2 \phi^{4}(z) } 
\,.
\label{eq:adiabatic-potentials-0}
\end{eqnarray}

%The analogy is not perfect, since
%the diamagnetic potential $\propto B^2$ does not appear here.
A formulation that is slightly more explicit than Eq.\,(\ref{eq:BdG-adiabatic-basis-0})
is the following second-order system 
for the adiabatic phase and density modes~\citep{Diallo15a}
\begin{eqnarray}
- \frac{ {\rm d}^2 \uTilde }{ {\rm d}z^2 } 
+ U_{\rm ad} \, \uTilde = \hat{L} \vTilde
\,,
&& \qquad
U_{\rm ad} = M_{\uTilde} + (\theta'/2)^2
\,,
\label{eq:BdG-adiabatic-phi}
\\[2ex]
- \frac{ {\rm d}^2 \vTilde }{ {\rm d}z^2 } 
+ V_{\rm ad} \, \vTilde = - \hat{L} \, \uTilde
\,,
&& \qquad
V_{\rm ad} = M_{\vTilde} + (\theta'/2)^2
\,.
\label{eq:BdG-adiabatic-f}
\end{eqnarray}
The derivatives of the condensate density appear in the potentials and, 
via the differential operator
\begin{equation}
\hat{L} = 
\frac{\theta''}{2} 
+ 
\theta' \frac{ {\rm d} }{ {\rm d}z } 
\,,
\label{eq:def-L}
\end{equation}
they also couple the amplitudes $\uTilde, \vTilde.$
The decoupling of the two Bogoliubov modes has thus nearly succeeded
in the adiabatic basis: 
if the variation of the condensate density is sufficiently slow or 
the energy large enough,
we expect the “source terms” $\hat{L} \, \vTilde$, $-\hat{L} \, \uTilde$
on the rhs of Eqs.\,(\ref{eq:BdG-adiabatic-phi}, \ref{eq:BdG-adiabatic-f}) to be small.

\begin{figure}[htbp]
   \centerline{%
\includegraphics*[height=0.42\textwidth]{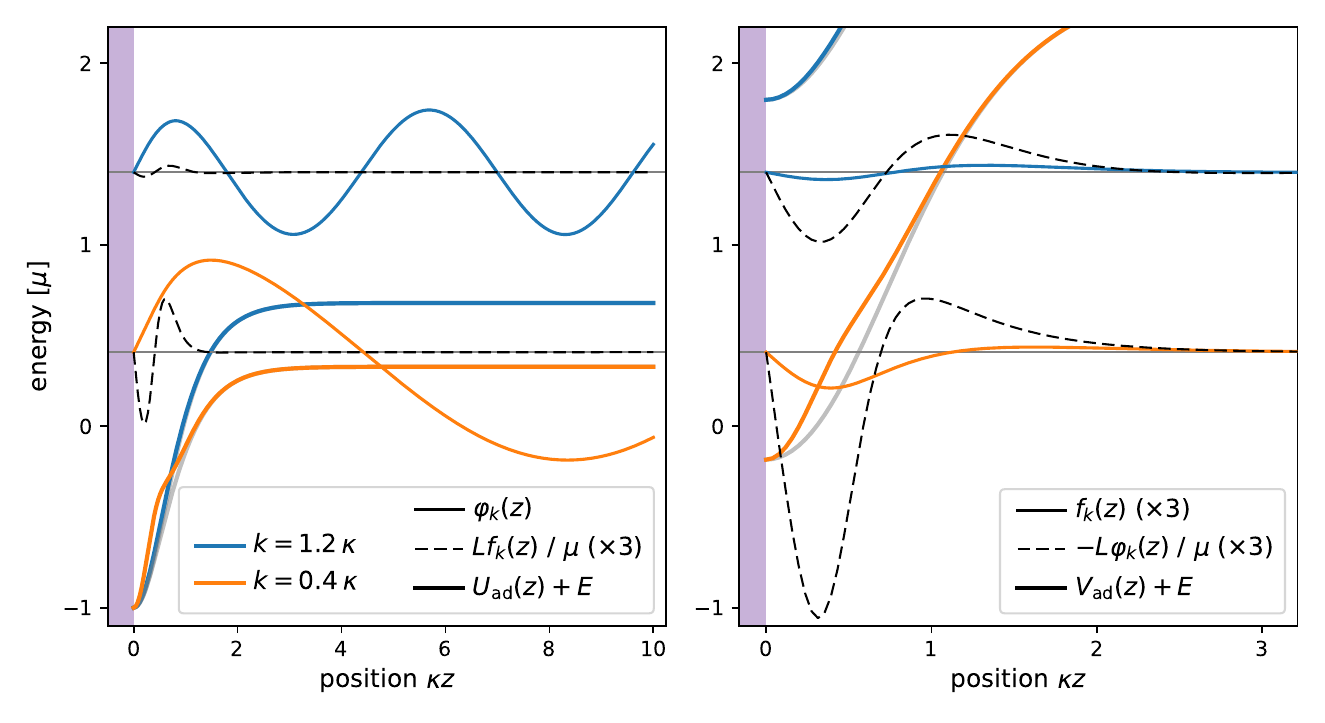}%
}
\caption[]{%
Bogoliubov mode functions $\uTilde$ (left) and $\vTilde$ (right)
for the hard-wall potential in the adiabatic representation. 
For clarity, we have shifted the potentials (thick solid lines)
of Eqs.\,(\ref{eq:BdG-adiabatic-phi}, \ref{eq:BdG-adiabatic-f})
by the energy $E$ to $U_{\rm ad}(z) + E$ (left) and $V_{\rm ad}(z) + E$ (right);
the mode functions are plotted on these
“energy levels” $E \approx 1.4\,\mu$ (blue), and $0.41\,\mu$ (orange).
The $k$-vectors (see inset legend) are computed 
from the dispersion relation~(\ref{eq:Bogoliubov-dispersion}), 
and $\kappa = \sqrt{\mu/2}$.
The black dashed curves illustrate the “source terms” 
$\hat{L} \vTilde$ (left)
and
$-\hat{L} \, \uTilde$ (right)
in Eqs.\,(\ref{eq:BdG-adiabatic-phi}, \ref{eq:BdG-adiabatic-f}).
To enhance visibility, a few curves are multiplied as indicated in 
the legends. The thick gray curves correspond to the eigenvalues
$M_{\uTilde}(z)$, $M_{\vTilde}(z)$ of 
Eq.\,(\ref{eq:adiabatic-potentials-0}), the bumps in the coloured
curves are due to the “geometric” potential $(\theta'(z)/2)^2$ in
Eqs.\,(\ref{eq:BdG-adiabatic-phi}, \ref{eq:BdG-adiabatic-f}).
}
   \label{fig:uv_adiabatic}
\end{figure}
%
% $L$ translates the inhomogeneity 
%of the condensate.
%• 
%Explain equations and fix words/concepts. 
%• 

This is even true for the hard-wall potential, 
as illustrated in Fig.\,\ref{fig:uv_adiabatic} where we plot
two adiabatic mode functions $\uTilde_k$, $\vTilde_k$, 
computed from the exact solution~(\ref{eq:uv-hard-wall}).
The function $\uTilde_k$ appears “above the barrier” $U_{\rm ad}$, oscillates
asymptotically with a constant wavenumber $k$, and merges
into the asymptote~(\ref{eq:asymptotic-u-and-v-0}) when the
rotation~(\ref{eq:rotated-basis}) is undone.
The opposite is true for $\vTilde_k$ 
which only appears due to the non-adiabatic coupling:
it is therefore localised close to the condensate boundary and vanishes
deep into the bulk.
We have checked that $\vTilde_k$ lies \emph{below} the eigenfunctions in the potential 
well $V_{\rm ad}$ subject to the boundary condition at $z = 0$.
%(see Sec.\,\ref{s:Feshbach} below).
Note also that the modes and their source terms scale quite differently:
the phase mode $\uTilde_k$ is much larger than $\hat{L} \, \vTilde_k / \mu$,
while the density mode $\vTilde_k$ is comparable to $-\hat{L} \, \uTilde_k / \mu$.

It is instructive to work through this formulation with the 
local density approximation (LDA). 
The latter amounts to replacing the chemical potential 
$\mu \mapsto \mu(z) = \mu - V(z)$ 
and to use the Bogoliubov spectrum. 
Inverting Eq.\,(\ref{eq:Bogoliubov-dispersion}), 
this leads to a local $k$-vector with
\begin{equation}
% 0 = k^4 + k^2 2 mu - E^2 
% k^2 = V(z) - mu + sqrt( [mu - V(z)]^2 + E^2)
\mbox{LDA:}\quad k^2(z) = V(z) - \mu + \left[ (\mu - V(z))^2 + E^2 \right]^{1/2}
\label{eq:LDA-k-of-E}
\end{equation}
If the Thomas-Fermi approximation is accurate, we have
$V_2(z) \approx g \phi^2(z) \approx \mu - V(z)$ and 
$k^2(z) \approx -M_{\uTilde}(z)$ 
from Eq.\,(\ref{eq:adiabatic-potentials-0}).
The adiabatic scheme goes beyond the LDA by including 
corrections to
the Thomas-Fermi approximation and gradients in the
condensate density (derivative $\theta'$ and operator $\hat{L}$).

\subsection{Adiabatic approximation}

The data of Fig.\,\ref{fig:uv_adiabatic}
suggest an approximate solution to the Bogoliubov problem 
that we call “adiabatic” in the following.
In a first step, we neglect the small source term $\hat{L} \, \vTilde$
in Eq.\,(\ref{eq:BdG-adiabatic-phi}) and solve
\begin{equation}
- \frac{ {\rm d}^2 \uTilde^{({\rm ad})} }{ {\rm d}z^2 } 
+ U_{\rm ad} \, \uTilde^{({\rm ad})} = 0
%\,,\qquad
%U_{\rm ad} = M_{\uTilde} + (\theta'(z)/2)^2
\label{eq:adiabatic-for-u}
\end{equation}
In the second step, this result is used as an approximation to the source term
$\hat{L}\uTilde$ in Eq.\,(\ref{eq:BdG-adiabatic-f})
to yield $\vTilde^{({\rm ad})}.$

The potential $U_{\rm ad}$ in the Schrödinger equation~(\ref{eq:adiabatic-for-u})
typically has a single turning point around $z \approx 0$, 
as illustrated in Fig.\,\ref{fig:split_U_ad}.
We re-write it in the form
\begin{equation}
U_{\rm ad}(z) = V_1(z) + D(z) + (\theta'/2)^2
\label{eq:split_U_ad}
\end{equation}
where the first term $V_1(z) = V(z) - \mu + g \phi^2(z)$ was encountered 
in Eq.\,(\ref{eq:def-H123}).
The second piece
\begin{equation}
D(z) 
= g \phi^2(z) - \big[ E^2 + g^2 \phi^4(z) \big]^{1/2}
= 
- \frac{ E^2 }{ g \phi^2(z) + \big[ E^2 + g^2 \phi^4(z) \big]^{1/2} }
\label{eq:def-R}
\end{equation}
monotonically increases from $-E$ to zero,
as the condensate density grows. 
It determines the long-range behaviour of $U_{\rm ad}$: 
the approach to its asymptote is slower for large $E$,
as can be seen in Fig.\,\ref{fig:split_U_ad}.
Finally, the term $(\theta'/2)^2 = V_{\rm geo}$ in 
Eq.\,(\ref{eq:split_U_ad}) 
has been called the “geometric” potential \citep{Diallo15a}.
Similar to the geometric phase,
it arises because the adiabatic basis is position-dependent.
$V_{\rm geo}$ is peaking in the transition region around $z = 0$
where the gradient in the condensate density is particularly large
(compare thick coloured and thick gray curves in Fig.\,\ref{fig:uv_adiabatic},
see also Fig.\,\ref{fig:split_U_ad} in the~\ref{a:plots}).
Broadly speaking, we find the hard-wall (linear) potential to give the largest
(smallest) geometric potential, respectively, but this depends, of course,
on the specific values of the parameters ($\kappa$ and $a$).
We come back in Sec.\,\ref{s:linear-phase}
to the linear case where we have to address 
the slow spatial decay of $U_{\rm ad}(z; E)$.

Numerical solutions within the adiabatic approximation are easy to find 
for any energy $E$
because the exponentially growing “contamination” has been eliminated.
A simple initial value solver suffices,
starting in the tunnelling region left of the turning point. 
This gives the phase mode
$\uTilde^{({\rm ad})}$ from which we compute the source
term $-\hat{L} \uTilde^{({\rm ad})}$ in Eq.\,(\ref{eq:BdG-adiabatic-f}).
%\begin{equation}
%- \frac{ {\rm d}^2 \vTilde }{ {\rm d}z^2 } + V_{\rm ad}\, \vTilde
%= - \hat{L}\,\uTilde
%\label{eq:BdG-adiabatic}
%\end{equation}
To solve the inhomogeneous Schrödinger equation for $\vTilde^{({\rm ad})}$,
we go back to a finite-difference or finite-element scheme, 
using a reasonably large interval.
Since the sought solution 
$\vTilde^{({\rm ad})}$ is below the spectrum of the 
potential $V_{\rm ad} = M_{\vTilde} + V_{\rm geo}$ and the source term
is spatially localised, 
there is no problem with inverting the Schrödinger operator 
in Eq.\,(\ref{eq:BdG-adiabatic-f}) on a finite grid.
% • • • 
% well, tunnelling tails arise
% • • • 

%The overall features of the adiabatic scheme
%can be understood from an approximate
%solution to Eq.\,(\ref{eq:BdG-adiabatic}), second line. 
%In the spirit of the Thomas-Fermi solution, neglect the
%second derivative on the lhs and get
%\begin{equation}
%\vTilde^{({\rm ad})}(z) \approx - \frac{ (\hat{L} \, \uTilde)(z) }{ V_{\rm ad}(z) }
%%\,,\qquad
%%V_{\rm ad}(z) = M_{\vTilde}(z) + [\theta'(z)/2]^2
%\label{eq:adiabatic-solution-tilde-v}
%\end{equation}
%In the spatial range where $\uTilde$ oscillates, this is small
%because the potential $V_{\rm ad}(z)$ is large (see Fig.\,\ref{fig:uv_adiabatic})
%and because the differential operator $\hat{L}$ 
%[see Eq.\,(\ref{eq:def-L})] localises the numerator
%to the region where the condensate gradient is nonzero.

\begin{figure}[htbp]
\centerline{%
\includegraphics*[height=0.42\textwidth]{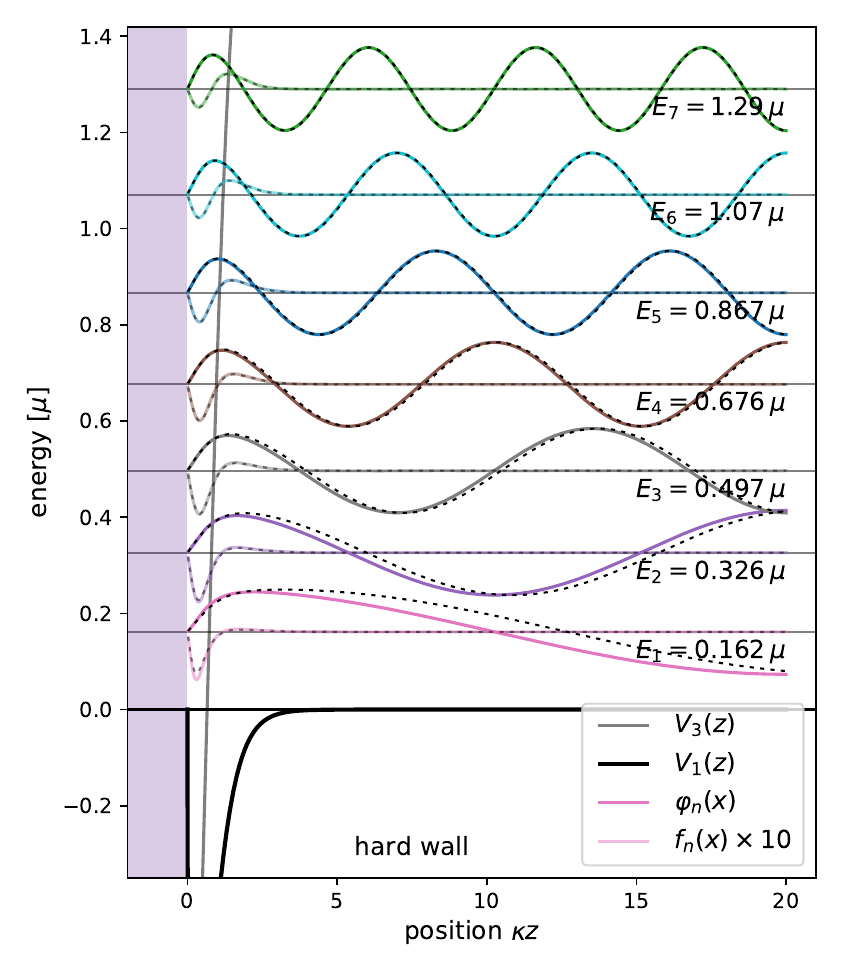}%
\hspace*{3mm}%
\includegraphics*[height=0.42\textwidth]{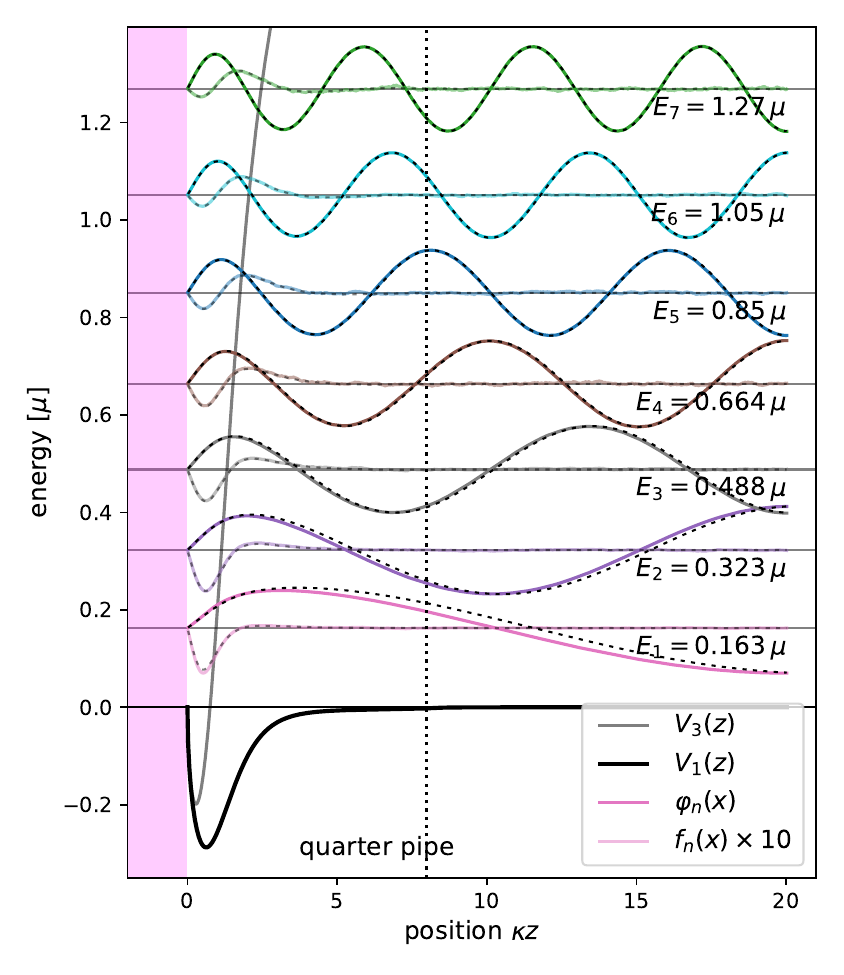}}

\centerline{%
\includegraphics*[height=0.42\textwidth]{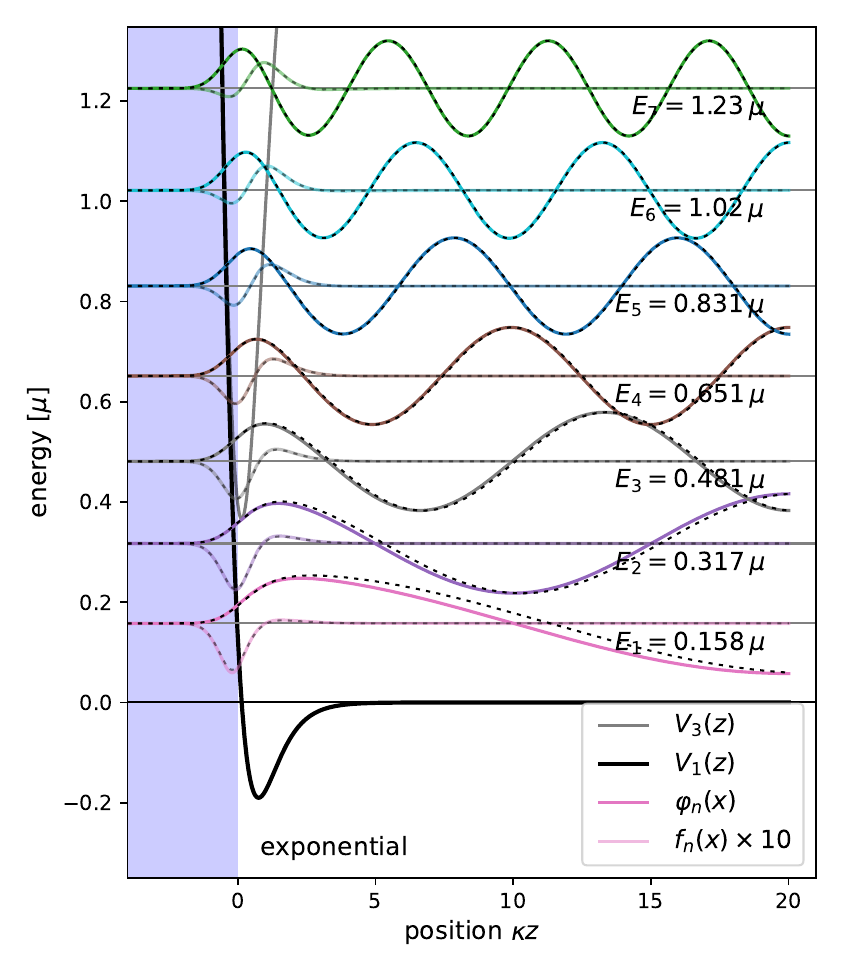}%
\hspace*{3mm}%
\includegraphics*[height=0.42\textwidth]{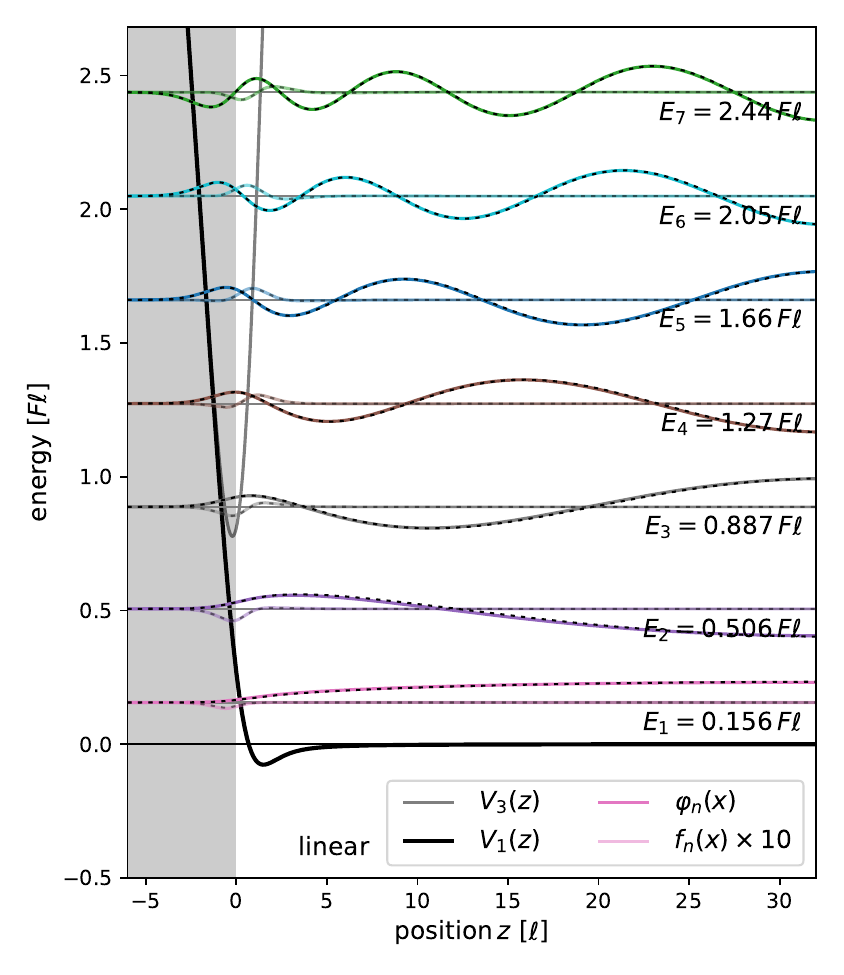}}

\caption[]{Comparison of exact (numerical) and adiabatic solutions
for the Bogoliubov modes in the density/phase ($\uTilde, \vTilde$) 
representation.
Exact results in color, adiabatic solutions black dotted.
Numerical parameters as in Table~\ref{t:units},
except $\kappa R = 8$ for the quarter-pipe case (vertical dotted line).
The plotted energies are somewhat arbitrary, as they are determined 
by boundary conditions at the right end of the computational grid.
Here, a Neumann condition is applied at $z \approx 20/\kappa\,(32\,\ell)$ 
for flat-bottom (linear) potentials, respectively.
}
\label{fig:adiabatic-phif}
\end{figure}

The results of this procedure are illustrated for a few energies in 
Fig.\,\ref{fig:adiabatic-phif}.
The dotted curves (mostly superimposed) correspond to the adiabatic
approximation, while
the coloured curves give numerically computed solutions of the Bogoliubov
equations with an (artificial) Neumann boundary condition.
Note the different character and magnitude of the $\uTilde$ and $\vTilde$ 
amplitudes, as expected from the above discussion.
The performance of the adiabatic scheme is fairly good, 
except at low energies.
In this case, one indeed expects from the definition~(\ref{eq:def-adiabatic-angle})
that the rotation angle $\theta(z)$ varies more rapidly with position, 
invalidating the adiabatic approximation.

\subsection{Reflection phase shift}
\label{s:phase-shift-1}

At this point, we address the main focus of this study, namely the phase
shift of the Bogoliubov waves at a generic turning point. 
We define it, as illustrated in Fig.\,\ref{fig:adiabatic-phif},
% Eqs.\,(\ref{eq:uv-hard-wall}, \ref{eq:delta-hw}) 
from the oscillatory $\uTilde$ solution in the adiabatic basis.
At large distance from the turning point, 
the density mode $\vTilde$ becomes small, 
and $\uTilde$ solves a Schrödinger equation 
with a relatively simple potential.
The solutions in this potential, analytically known, 
have been given in 
Eqs.\,(\ref{eq:uv-hard-wall}) 
and~(\ref{eq:Bessel-Coulomb-solutions-0});
they
provide reference waves, and the phase shift $\delta$ is measured
relative to them.

% Provide reference potential/wave solution and define phase shift.

In the flat-bottom potentials of type “hard wall”,
“quarter pipe” and “exponential”, the condensate becomes constant,
$\phi^2(z) \to \mu / g$,
when the distance $z$ is large enough compared to the scales $1/\kappa$ or $1/a$.
The non-adiabatic terms proportional to $\theta'$ vanish, and
in the potential $U_{\rm ad}$ [Eq.\,(\ref{eq:split_U_ad})], only the term
$D(z)$ survives. 
It tends to the constant value
\begin{equation}
U_{\rm ad}(z) 
\to 
%\mu - (E^2 + \mu^2)^{1/2}
  - \frac{ E^2 }{ \mu + (E^2 + \mu^2)^{1/2} }
= - k^2
\label{eq:limiting-value-U_ad}
\end{equation}
which is the wavenumber $k$ of the Bogoliubov dispersion~(\ref{eq:Bogoliubov-dispersion}), as expected from the local density approximation.
Asymptotically, the solution $\uTilde(z)$ is therefore 
a combination of sine and cosine waves.
Similar to Eq.\,(\ref{eq:def-phase-2}),
we define the phase shift $\delta$ by the limit
\begin{equation}
z \to \infty: \quad
\uTilde(z) \to
\frac{ A }{ \sqrt{k} } \sin( k z + \delta ) 
= A \left[ 
  j_k(z) \cos \delta
- y_k(z) \sin \delta
\right]
\label{eq:def-phase-1}
\end{equation}
with an amplitude $A$.
This is consistent with Eqs.\,(\ref{eq:asymptotic-u-and-v-0}) for the
standard Bogoliubov amplitudes $u$, $v$ because the adiabatic angle
becomes constant, $\theta(z) \to \arctan( \mu / E ) = \theta_k$ 
{}[Eqs.\,(\ref{eq:theta-angle-hw}, \ref{eq:def-adiabatic-angle})].

The pair of “reference waves” 
$\{ j_k, y_k \} = k^{-1/2} \left\{ \sin(kz), -\cos(kz) \right\}$ 
in Eq.\,(\ref{eq:def-phase-1}) are normalised 
such that their Wronskian $W[j_k, y_k]$ is unity.
The convention ensures that incident or reflected waves given by 
$j_k \mp {\rm i} y_k$ have unit flux.
This is based on the flux of linear wave mechanics, although for the actual 
particle current in a Bose gas, the interference with the condensate has 
to be taken into account \citep{Bouchoule03, Paul05a}.
As mentioned around its definition in Eq.\,(\ref{eq:def-Wronskian}),
the Wronskian is a convenient tool to extract the phase shift $\delta(E)$
of the $\uTilde$ solution. Indeed, from Eq.\,(\ref{eq:def-phase-1})
%•
%signs double-checked
%•
\begin{equation}
%W[\uTilde^{({\rm ad})}, j_k] \to A \, \sin\delta^{({\rm ad})}
%\,,\qquad
%W[\uTilde^{({\rm ad})}, y_k] \to A \, \cos\delta^{({\rm ad})}
z \to \infty: \quad
W[\uTilde, j_k] \to A \, \sin\delta
\,,\qquad
W[\uTilde, y_k] \to A \, \cos\delta
\label{eq:get-phase-from-W}
\end{equation}
The derivative of the Wronskian is 
\begin{eqnarray}
\frac{ {\rm d} }{ {\rm d}z } W[\uTilde, j_k] &=& 
\uTilde \frac{ {\rm d}^2 j_k }{ {\rm d}z^2 } 
-
\frac{ {\rm d}^2 \uTilde }{ {\rm d}z^2 } j_k
\nonumber\\
&=&
- \uTilde \, k^2 j_k
+
\left( - U_{\rm ad} \uTilde + \hat{L} \vTilde
\right) 
j_k 
\nonumber\\
&=&
- \uTilde \left( U_{\rm ad} + k^2 \right) j_k
+
(\hat{L} \vTilde) \, j_k
\label{eq:Abel-theorem-0}
\end{eqnarray}
The last term arises from the rhs of the inhomogeneous 
Schrödinger equation for $\uTilde$.
Its contribution to $\delta$ is evaluated in Sec.\,\ref{s:phase-shift-2}.
Otherwise, we find the difference between
the actual potential $U_{\rm ad}$ and the constant 
reference $-k^2$ of Eq.\,(\ref{eq:limiting-value-U_ad}),
a result known as the Abel identiy \citep{BenderOrszag}.
At large $z$, both terms in Eq.\,(\ref{eq:Abel-theorem-0}) vanish, 
and $W$ becomes constant,
as anticipated in Eq.\,(\ref{eq:get-phase-from-W}).
%The limiting value defining the phase is taken in the region
%$\kappa z \gg 1$ (hard wall), 
%$z > R$ (quarter pipe),
%or 
%$a z \gg 1$ (exponential potential),
%respectively, 
We use this constant value to normalise the solution
$\uTilde$ (and $\vTilde$) such that $A = 1$ in Eq.\,(\ref{eq:def-phase-1}).

\begin{figure}[tbh]
\centerline{%
\includegraphics*[height=0.3\textwidth]{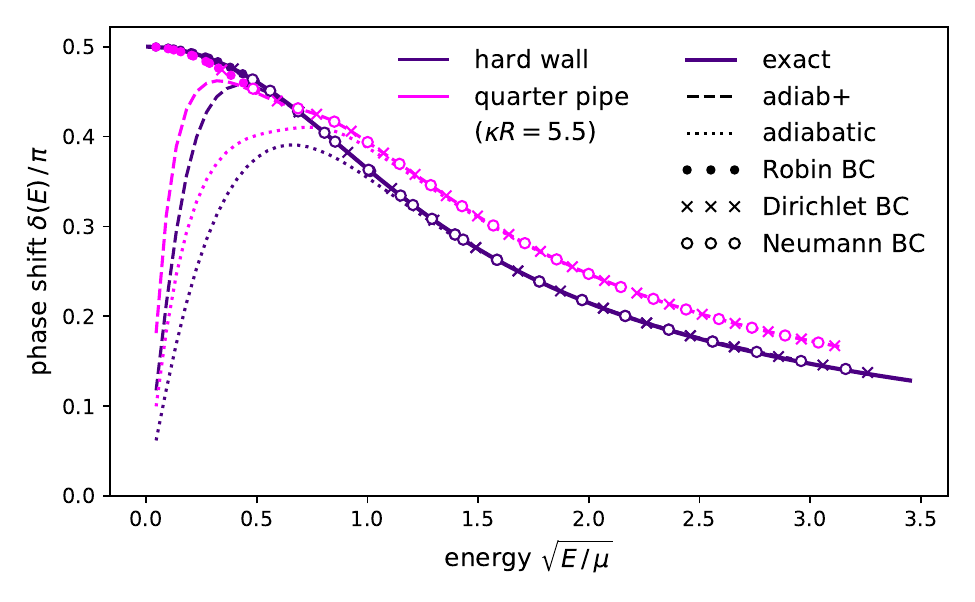}%
\hspace*{3mm}%
\includegraphics*[height=0.3\textwidth]{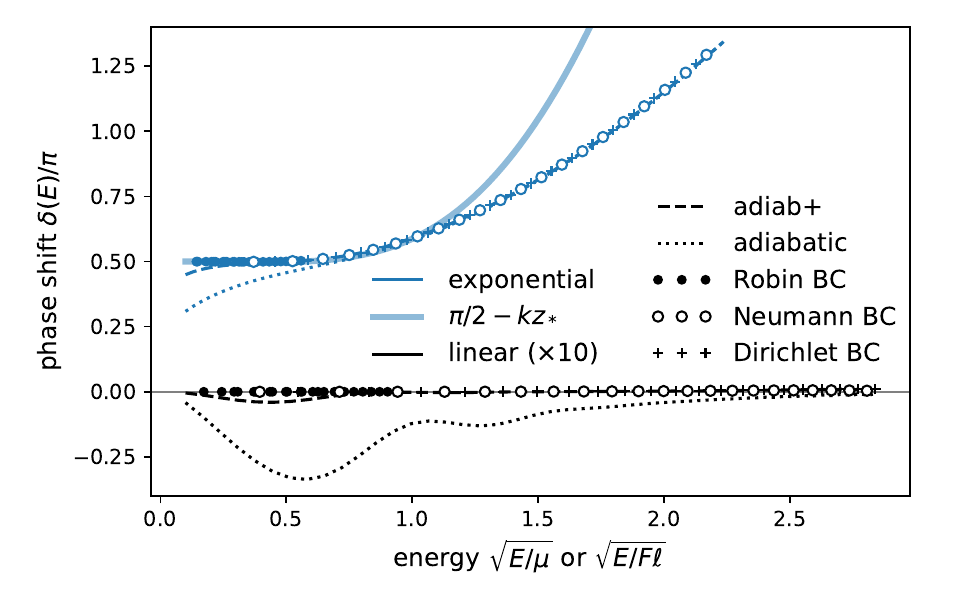}%
}
\caption[]{%
%• • • 
%very low energy solutions and their phase shifts to add 
%• • •
Phase shift $\delta(E)$ of the oscillatory Bogoliubov mode $\uTilde(z)$,
obtained by matching it to the asymptotic form~(\ref{eq:def-phase-1})
or~(\ref{eq:def-phase-2}) (for the linear potential).
Symbols: numerical diagonalisation with
boundary condition at the open right end as indicated,
dashed lines (“adiab+”): adiabatic approximation plus the 
non-adiabatic correction discussed in Sec.\,\ref{s:phase-shift-2}, 
dotted lines (“adiabatic”): without this correction.
Thick solid line (left panel): exact solution Eq.\,(\ref{eq:delta-hw});
pale thick solid line (right panel): simple turning-point construction
$\delta = \pi/2 - k z_*$ (see main text).
In the linear potential (right panel), the phase shift is extrapolated
to infinity (Sec.\,\ref{s:linear-phase}), 
starting from the position $z \approx 32\,\ell$ 
(see Table~\ref{t:units} for the units). 
Parameters: quarter-pipe potential with $\kappa R = 5.5$,
exponential potential with $\kappa / a = 1/\sqrt{2}$.
We plot against the root of the mode energy to enlarge
the low-energy region. 
Results for the quarter-pipe potential with other parameters
are shown in Fig.\:\ref{fig:scan-R}.
}
\label{fig:phase-shift-1}
\end{figure}

The phase shift $\delta = \delta(E)$ quantifies to what extent the
inhomogeneous condensate background “drags” or “pushes” the Bogoliubov wave 
into or out of the region near the turning point. 
% • positive or negative $\delta$ / check sign •
% OK: positive delta means shifting the sin reference wave to the left,
% 'into' the turning point region.
This can be understood in terms of the dip
of the effective potential $U_{\rm ad}(z)$ in Fig.\,\ref{fig:split_U_ad}:
the local wavelength of the $\uTilde$ wave decreases, and more phase
is accumulated relative to a flat potential (positive $\delta$).
%•
%From the plots in Fig.\,\ref{fig:split_U_ad}, a much larger phase is
%expected (think of the integral over $U_{\rm ad}$ with the asymptote
%subtracted).
%WKB calculation with Thomas-Fermi approximation, keeping
%only the $D(z)$ contribution? ... just for the trend -- $D(z)$ alone
%is roughly one half of the phase shift. The agreement with the WKB
%calculation is quite OK, semi-quantitative for the exponential potential.
%The pi/4 Langer phase is essential, it does not appear for the half-pipe
%potential (at least the error is smaller than that).
%•
The four potentials show quite some diversity in their phase shifts,
as can be seen in Fig.\,\ref{fig:phase-shift-1}.
The exact result in the hard-wall potential (solid line) is reproduced 
by the numerical solution, while the adiabatic approximation 
fails at low energies only (dashed lines).
The non-adiabatic correction discussed in Sec.\,\ref{s:phase-shift-2}
below
is also quite significant at low energies. It is represented 
by the difference between the dotted and dashed lines.
This applies to all types of potentials.

The energy dependence of $\delta(E)$ makes it plausible why 
the level spacings in Fig.\,\ref{fig:adiabatic-phif} slightly
differ between the three flat-bottom potentials.
For the exponential potential, the larger phase shift moves
the levels down (smaller energy spacing for a fixed right end of the 
computational grid),
compared to the hard-wall potentials
whose phase shift decreases with energy.
This trend correlates with the ordering of the energy levels, e.g.,
$E_3 = 0.497\,\mu > 0.488\,\mu > 0.481\,\mu$, when going from top left
to bottom left in Fig.\,\ref{fig:adiabatic-phif}, even though close
to their turning points, the Bogoliubov modes are strikingly similar.

%We have compared one by one the contributions to $\delta(E)$ from
%the three terms in $U_{\rm ad}$ [Eq.\,(\ref{eq:split_U_ad})], and
%• • • different trends, non-trivial compensations? • • •

%The same trend also holds when comparing to the numerically exact
%solutions (symbols in Fig.\,\ref{fig:phase-shift-1}) for the other potentials.
%We recall that the low-energy limit can be handled exactly because
%Eqs.\,(\ref{eq:BdG-phase}) decouple with the solutions $\varphi \to \phi$
%(the condensate) and $f \to 0$ (under the $V_3$ barrier).
%We then expand
%\begin{equation}
%E \to 0: \qquad
%\uTilde(z; E) \to \frac{ \sin \delta + k z \cos \delta + \ldots }{ \sqrt{k} }
%\label{eq:}
%\end{equation}
%This matches with the constant value of the condensate $\phi$ 
%only for $\cos\delta = 0$, i.e., $\delta = \pm \pi/2$. 

The phase shift in the quarter-pipe potential is similar to the
hard wall, with a slightly larger value for most energies
[Fig.\,\ref{fig:phase-shift-1}(left)].
We attribute this to a somewhat larger region (of the order of the
quarter-pipe radius $R$) where the condensate
density deviates from its constant asymptote so that $U_{\rm ad}(z)$
develops a deeper dip;
compare the top panels in Fig.\,\ref{fig:sketches} 
and Fig.\,\ref{fig:scan-R} in \ref{a:plots}.
In the smooth exponential potential, the phase shift starts from
the same low-energy value $\delta(0) \approx \pi/2$ as for the
hard wall, but then increases monotonously with energy
[Fig.\,\ref{fig:phase-shift-1}~(right)].
We attribute this to the “softness” of the potential $U_{\rm ad}$ 
that provides additional “space” for the wave function
in the region $z < 0$.
Indeed, the turning point $z_*(E)$,
the root of $U_{\rm ad}(z_*) = 0$,
moves to the left as $E$ increases.
Crudely speaking, one may imagine the Bogoliubov wave to be simply 
translated from the reference point $z = 0$ to its actual turning point 
$z_*$ \citep{Henkel94a}.
This picture leads to the simple approximation $\delta = \pi/2
- k z_*$ which captures well the low-energy behaviour
[thick solid line in Fig.\,\ref{fig:phase-shift-1}(right)].
We have also computed within the semiclassical (WKB) approximation 
\citep{BenderOrszag}
the phase of
the $\uTilde^{({\rm ad})}$ wave function 
and found good agreement.
It is somewhat remarkable that the WKB phase correction $\pi/4$
at the soft turning point in the exponential potential accounts
only for one half of the low-energy limit $\delta(0) = \pi/2$.
At low energies, of course, the semiclassical approximation breaks down 
because of the rapid variation of the local momentum. 
% see Fig.\,\ref{fig:adiabatic-phif}.

In contrast, the linear potential produces a phase shift close to zero 
(note that the data in Fig.\,\ref{fig:phase-shift-1}, right are
magnified for better visibility). 
These small values are found once the phase shift is extrapolated 
from the right end point of the numerical grid to infinity:
this is discussed in more detail in Sec.\,\ref{s:linear-phase}.
Before, we address the correction to $\delta(E)$ arising from the
term $\hat{L} \, \vTilde$ neglected so far
in the adiabatic approximation.
Including it moves the dotted curves to the dashed ones 
and improves the agreement with the exact data in the range $0.25\,\mu < E < \mu$
(and even for all $E$ in the linear potential).

\subsection{Non-adiabatic correction to phase shift}
\label{s:phase-shift-2}
%Develop theory for phase shift (Wronskians à la Messiah) and include
%the back-action of the density mode (integro-differential equation,
%not really Feshbach-like because no strong resonance.

%In Fig.\,\ref{fig:phase-shift-1}, 
%small deviations were visible
%in low-energy phase shift, comparing the numerically exact calculation 
%of $\uTilde$ and its adiabatic approximation $\uTilde^{({\rm ad})}$.
%•
%Not seen in the plot.
%•
%The disagreement is resolved in this subsection.

The adiabatic solution $\uTilde^{({\rm ad})}$ deviates from the 
full one $\uTilde$ 
because it neglects the inhomogeneous term $\hat{L} \, \vTilde$
in Eq.\,(\ref{eq:BdG-adiabatic-phi})
and in Eq.\,(\ref{eq:Abel-theorem-0}).
We use again the Abel identity to improve,
if not the phase mode, but at least its phase shift.
The technique is well known in partial wave expansions 
of scattering theory \citep{MessiahI}.
Consider therefore the Wronskian $W[\uTilde, \uTilde^{({\rm ad})}]$ 
between phase modes defined as in Eq.\,(\ref{eq:def-Wronskian}).
Computing its derivative and using the Schrödinger-type equations
for $\uTilde$, $\uTilde^{({\rm ad})}$ at the same energy,
we find [similar to Eq.\,(\ref{eq:Abel-theorem-0})]
\begin{equation}
\frac{ {\rm d} }{ {\rm d}z } W[\uTilde, \uTilde^{({\rm ad})}] 
= 
% with other sign convention for W and diff eqn for u: plus sign
(\hat{L} \, \vTilde)(z) \,
\uTilde^{({\rm ad})}(z)
\label{eq:diff-for-W-and-Lv}
\end{equation}
Recall from Fig.\,\ref{fig:uv_adiabatic}
that the term $\hat{L} \, \vTilde$ is relatively small
and well localised.
As we move infinitely far away from the condensate border, 
the expansion of $\uTilde$ and $\uTilde^{({\rm ad})}$ in reference 
solutions 
yields for their Wronskian (assuming for both the normalisation $A = 1$)
%•
%sign double-checked:
%% W[j, y] = 1 and
%% W[j cos d - y sin d, j cos a - y sin a] = 
%% - cos d sin a + sin d cos a = sin(d - a)
%•
\begin{equation}
z \to \infty: \quad
W[\uTilde, \uTilde^{({\rm ad})}] 
%= 
%- \cos \delta \, \sin \delta^{({\rm ad})}+ \sin \delta \, \cos \delta^{({\rm ad})}
= \sin(\delta-\delta^{({\rm ad})})
\label{eq:}
\end{equation}
At the other end of the $z$-axis, $W$ drops to zero, 
since $\uTilde$ and $\uTilde^{({\rm ad})}$
vanish as physical tunnelling solutions.
By integrating Eq.\,(\ref{eq:diff-for-W-and-Lv}), we thus get the 
exact expression
\begin{equation}
\sin(\delta-\delta^{({\rm ad})}) 
= 
\int\!{\rm d}z \, 
(\hat{L} \, \vTilde)(z) \,
\uTilde^{({\rm ad})}(z)
=
- \int\!{\rm d}z \, 
\vTilde(z) \,
(\hat{L} \, \uTilde^{({\rm ad})})(z)
\label{eq:phase-shift-corrn-and-Lv-u-integral}
\end{equation}
The second form arises from a partial integration, using the 
definition~(\ref{eq:def-L}) of the differential operator $\hat{L}$.
The integrated terms vanish because the integrand is localised
around the turning point region of the condensate.
This expression can be re-arranged into (incidentally another
Wronskian)
\begin{equation}
\sin(\delta-\delta^{({\rm ad})}) 
=
\frac12
\int\!{\rm d}z \,
\theta'(z)
\left[
\vTilde'(z) \,
\uTilde^{({\rm ad})}(z)
-
\vTilde(z) \,
\uTilde^{({\rm ad})\prime }(z)
\right]
\label{eq:}
\end{equation}
where the prime denotes spatial derivatives.
We found this form to be numerically convenient 
because the integrand shows less oscillations.
%• • • 
%add a figure in Appendix~\ref{a:plots}
% oscillatory features do not look very nasty ...
%• • • 

At this point, it is instructive to invert
the Schrödinger operator $-{\rm d}^2 / {\rm d}z^2 + V_{\rm ad}(z)$
with the help of the resolvent operator $\hat{G}$.
As illustrated in Fig.\:\ref{fig:uv_adiabatic}, the potential $V_{\rm ad}(z)$
is a confining well: on the left side of the turning point, it is closed
by the bare trapping potential $V(z)$, on the right side, by the repulsive
mean-field potential $V_3(z)$.
The resolvent $\hat{G}$ provides the solution $\vTilde$ of
the inhomogeneous equation~(\ref{eq:BdG-adiabatic-f}),
which may be expanded as follows
\begin{equation}
\vTilde 
= - \hat{G} \hat{L} \, \uTilde 
= - \sum_{n=0}^{\infty} \eta_n \frac{ (\eta_n | \hat{L} \, \uTilde ) }{ \epsilon_n }
\label{eq:resolvent-for-v}
\end{equation}
Here, the $\epsilon_n$ and $\eta_n$ ($n = 0, 1, \ldots$) provide the discrete spectrum 
in the potential~$V_{\rm ad}$,
with eigenfunctions normalised according to the standard scalar product
\begin{equation}
(\eta_n | \eta_m) = \int\!{\rm d}z\, \eta_n(z) \,\eta_m(z) = \delta_{nm}
\,,\qquad
n,m = 0, 1, \ldots
\label{eq:v-nm-orthogonal}
\end{equation}
Since $V_{\rm ad}$ is a potential well, its eigenfunctions $\eta_n$ are localised. 
In particular for small quantum numbers $n$,
the matrix elements $(\eta_n | \hat{L} \, \uTilde )$ are convergent overlap 
integrals.

We insert the expansion~(\ref{eq:resolvent-for-v}) 
into the formula~(\ref{eq:phase-shift-corrn-and-Lv-u-integral})
for the non-adiabatic correction to the phase shift
and find
\begin{equation}
\sin(\delta-\delta^{({\rm ad})}) 
= 
\sum_n 
\frac{ (\eta_n | \hat{L} \, \uTilde^{({\rm ad})}) \,
(\eta_n | \hat{L} \, \uTilde ) }{ \epsilon_n }
\approx
\sum_n 
\frac{ |(\eta_n | \hat{L} \, \uTilde^{({\rm ad})})|^2
}{ \epsilon_n }
\label{eq:phase-shift-like-Feshbach}
\end{equation}
While the first expression is still exact, the second one is a useful
approximation. It is based on the assumption that the “back-action” of 
$\vTilde$ on the Bogoliubov mode $\uTilde$ is weak, as can be
expected from Fig.\,\ref{fig:uv_adiabatic}. 
% [• check norm of $\vTilde$ compared to ... •]
It also provides a way to correct the phase shift 
of the adiabatic approximation scheme 
where the density mode is
$\vTilde \approx \vTilde^{({\rm ad})} 
= - \hat{G} \hat{L} \uTilde^{({\rm ad})}$.
%[see Eq.\,(\ref{eq:resolvent-for-v})].
The results of this procedure are displayed as “adiab+” in 
Figs.\:\ref{fig:phase-shift-1}, \ref{fig:scan-R}.

The structure of Eq.\,(\ref{eq:phase-shift-like-Feshbach})
is reminiscent of second-order perturbation theory where the
eigenstates $\eta_n$ trapped in the potential $V_{\rm ad}$ are coupled 
to the wave function $\uTilde^{({\rm ad})}$ via the 
operator $\hat{L}$. 
A common feature of Eq.\,(\ref{eq:phase-shift-like-Feshbach}) 
and ground-state energy shifts is the definite sign: all terms
in the sum are positive, and
in distinction to a Feshbach resonance,
the energy denominator never crosses zero.
Still, there are oscillations in the “matrix element” 
that provide relatively broad peaks.
Indeed, as the energy $E$ is scanned,
the nodes and anti-nodes of $\uTilde^{({\rm ad})}$ shift across the bottom
of the potential $V_{\rm ad}(z)$ where the
low-lying eigenfunctions $\eta_n$ are localised.
%This behaviour is also at the origin of the non-monotonous energy dependence
%of the norm of the density mode $\vTilde$ plotted in Fig.\,\ref{fig:norm-v}.
%See \citet{Diallo15a} for a more detailed discussion in the linear potential.
These oscillations are particularly marked for the linear potential
which is the “softest” of the four,
compare dotted and dashed lines in Fig.\,\ref{fig:phase-shift-1}(right).
The turning point of $\uTilde$ is hence relatively free to move to 
the left.

\subsection{Phase shift in the linear potential}
\label{s:linear-phase}

%The “missing” $\pi/2$ at low energies can be explained by the
%different choice of the reference waves.

%Plot for model potentials.
%Challenge: long-range behaviour in linear potential, need to extrapolate
%phase shift (same technique).

%
%as discussed by us in earlier work \citep{Diallo15a}. 
%The reason is the linear increase of the condensate density
%$g \phi^2(z) \approx z$ [the leading order of 
%Eq.\,(\ref{eq:post-TF-condensate-density})].
%
%The WKB-like phase shift $+\pi/4$ only explains half of the difference
%in $\delta$ at low energies between the exponential and the linear potential,
%see Fig.\,\ref{fig:phase-shift-1}~(right).

The large-$z$ asymptotics in the linear potential 
has to be treated in a slightly different way,
because the approach of the potential $U_{\rm ad}$ to its
asymptotic form given in Eq.\,(\ref{eq:long-range-Coulomb-potential})
is slow.
As a consequence, when the parameters $A$, $\delta$ are computed 
at a finite position $z$ from the asymptotic form given in Eq.\,(\ref{eq:def-phase-2}),
they are still changing as $z$ moves to infinity. 
More explicitly, we compute $A \sin\delta$ from the Wronskian $W[\uTilde, j_E]$ 
with the regular reference solution $j_E$ [Eq.\,(\ref{eq:Bessel-Coulomb-solutions-0})].
Exactly as in Eq.\,(\ref{eq:Abel-theorem-0}),
we find that this Wronskian evolves according to
% the difference between the actual potential $U_{\rm ad}$ 
% and its long-range asymptote 
% (Abel theorem, \cite{BenderOrszag})
\begin{equation}
\frac{ {\rm d} }{ {\rm d}z } W[\uTilde, j_E] 
=
- \uTilde \, \Big(
  U_{\rm ad}
+ \frac{ E^2 }{ 2 z } 
+ \frac{ 1 }{ 4 z^2 } 
\Big) \,
j_E 
+
(\hat{L} \vTilde) \, j_E
\label{eq:slowly-evolving-sin-phase}
\end{equation}
using the natural units in the linear potential.
% \vspace*{1ex}
The last term involving $\hat{L} \vTilde$
has been dealt with in the previous section and because
it is localised to small $z$, it will
be neglected in the large-distance range considered here.
The three pieces of $U_{\rm ad} = V_1 + D + V_{\rm geo}$ introduced in Eq.\,(\ref{eq:split_U_ad})
can be expanded for large $z$.
Using the post-Thomas-Fermi expansion~(\ref{eq:post-TF-condensate-density}),
we get
\begin{eqnarray}
V_1(z) &=& - \frac{ 1 }{ 4 z^2 } - \frac{ 9 }{ 8 z^5 } 
+ {\cal O}(z^{-8})
\\
D(z) &=& 
- \frac{ E^2 }{ 2 z } 
+ \frac{ E^4 }{ 8 z^3 } 
- \frac{ E^2 }{ 8 z^4 } 
+ {\cal O}(E^6 z^{-5})
%- \frac{ E^6 }{ 16 z^5 } 
%+ {\cal O}(E^4 z^{-6})
\\
V_{\rm geo}(z) &=& \frac{ E^2 }{ 4 z^4 } + {\cal O}(E^4 z^{-6})
\label{eq:potential-asymptotes}
\end{eqnarray}
The corrections relative to the long-range 
reference potential of Eq.\,(\ref{eq:long-range-Coulomb-potential}) 
are thus
%[• sign checked (correction is positive!) •]
\begin{equation}
U_{\rm ad}(z) 
+ \frac{ E^2 }{ 2 z } 
+ \frac{ 1 }{ 4 z^2 } 
=
\frac{ E^4 }{ 8 z^3 }
+ \frac{ E^2 }{ 8 z^4 }
+ {\cal O}(z^{-5}, E^6 z^{-5})
%- \frac{ 18 + E^6 }{ 16 z^5 } 
%+ {\cal O}(E^4 z^{-6})
% full expression within Thomas-Fermi approximation, but no
% large-z simplification:
%  \frac{ E^4 / (4 z) }{ z^2 + z ( E^2 + z^2 )^{1/2} + E^2 / 2 }
\label{eq:leading-difference}
\end{equation}
%although higher terms decay only slightly faster.

\begin{figure}[bth]
\centerline{%
\includegraphics*[width=0.6\textwidth]{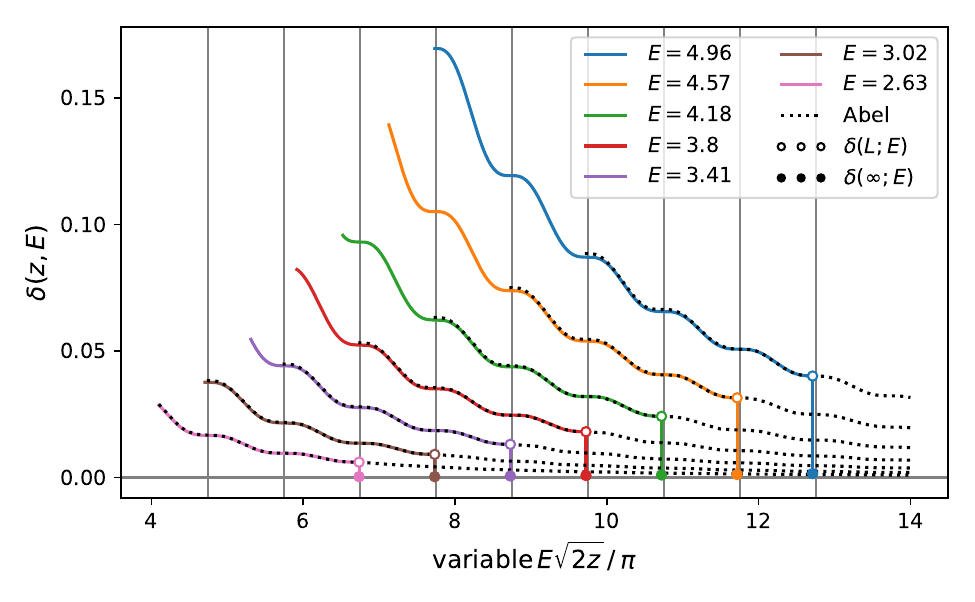}%
}
\caption[]{%
%• • • 
%adapt to showing phase directly, using the diff eqn~(\ref{eq:exact-d-delta-dz}).
%• • • 
Extrapolation of phase shift $\delta(z, E)$ to $z = \infty$.
The coloured curves give the numerically computed phase shift
based on solutions with Dirichlet boundary conditions at
$L \approx 32\,\ell$ (terminating at the open dots).
The vertical grid lines mark the zeroes of the reference solution $j_E( z )$.
The dotted black curves are obtained as explained after 
Eq.\,(\ref{eq:exact-d-delta-dz}),
starting from the open dots (phase shift at $z = L$) to both sides.
The phase shift resulting from continuing the dotted curves to infinity
is represented by the full dots with $\delta(E) \approx 0$.
}
\label{fig:extrapolate-W}
\end{figure}

Equation~(\ref{eq:slowly-evolving-sin-phase}) also provides
an equation for the phase shift $\delta$ alone. 
Consider amplitude $A$ and phase shift $\delta$ as the polar coordinates 
of the tuple
$(W[\uTilde, y_E]$, $W[\uTilde, j_E]) = (A \cos\delta, A \sin\delta)$. 
For the derivative of $\delta 
= \arctan( W[\uTilde, j_E] / W[\uTilde, y_E])$, 
a short calculation yields 
\begin{equation}
\frac{ {\rm d} \delta }{ {\rm d}z } \approx
-
\Big(
  \frac{ E^4 }{ 8 z^3 } 
+ \frac{ E^2 }{ 8 z^4 } 
\Big)
\left( j_E \cos \delta - y_E \sin \delta \right)^2
\qquad (z \gg \ell)
\label{eq:exact-d-delta-dz}
\end{equation}
At large distances, the last parenthesis can be simplified using
the asymptotic form~(\ref{eq:asymptotic-Bessel-Coulomb}) 
of the Bessel functions.
Eq.\,(\ref{eq:exact-d-delta-dz}) therefore yields an oscillatory
decrease of the phase shift,
as the position $z$ moves towards infinity.
If $\delta$ is already small, we have ${\rm d}\delta / {\rm d}z = 0$
at the nodes of the Bessel function $j_E(z) \propto J_0(E \sqrt{2z})$,
as illustrated in Fig.\,\ref{fig:extrapolate-W}.
This behaviour produces numerical artefacts when the phase shift
is computed at a finite position $z$. 
They show up as a slowly
increasing, oscillatory energy dependence.

%•
%Fig. to illustrate?
%•
%It turns out that one needs to integrate the differential equation
%to distances of the order $E$ to get a satisfactory estimate for $\delta$.
%This makes the numerical computation challenging at large energies.

Figure~\ref{fig:extrapolate-W} also illustrates the extrapolation
of the phase shift to infinity.
To find this value (filled dots),
we integrate Eq.\,(\ref{eq:exact-d-delta-dz}) from
the right end $z = L$ of the computational grid to convergence
near $z = \infty$.
%By repeated partial integrations, the result can be expressed in terms 
%of the cosine integral, but this is only marginally illuminating. 
The dotted curves permit to check the accuracy of the approximation
(the expansion of $U_{\rm ad}(z)$):
the extrapolation towards the left, $z < L$, works well.
We thus estimate phase shifts of the order $\delta \sim 10^{-4}\pi$,
probably consistent with zero within the accuracy of the numerical 
scheme.

%•
%Integral
%\begin{eqnarray}
%\sin\delta - 
%\sin\delta(z_1) &\approx 
%- E^3
%\int_{z_1}^\infty\!{\rm d}z \,
%\frac{ \sin^2( E \sqrt{2 z} + \pi/4 ) }{ (2 z)^{5/2} }
%\nonumber\\
%&=
%% x = E \sqrt{2 z}
%% dx = E dz / \sqrt{2 z}
%- E^6
%\int_{E \sqrt{2z_1}}^\infty\!{\rm d}x \,
%\frac{ \sin^2( x + \pi/4 ) }{ x^4 } % E^4 (2 z)^2 }
%\nonumber\\
%&= -\frac{ E^{6} }{ 6 x^3 } \left(
%4 x^3 \mathop{\rm Ci}(2 x)
%- 2 x^2 \sin(2 x) 
%+ x \cos(2 x) 
%+ \sin(2 x) + 1 
%\right)
%\nonumber\\
%&=
%- \frac{ E^{6} }{ 6 x^3 } % 12 x^5 }
%\left( 
%% 2 x^2 + 3 x \cos 2 x + 6 \sin 2 x
%1 + \frac{ 3 \cos 2 x }{ 2 x } + \frac{ 3 \sin 2 x }{ x^2 }
%+ \ldots
%\right)
%\label{eq:}
%\end{eqnarray}
%can be expressed with the cosine integral.
%• 
%Some crude first numerics does not agree with 
%Eq.\,(\ref{eq:slowly-evolving-sin-phase}).
%•

While our numerical results are subject to some finite error
margin, 
there are also analytical indications 
that the linear potential (Painlevé II problem) is a somewhat special case.
For example, the condensate density $|\phi(z)|^2$ is balanced exactly
when one integrates its “tunnelling tail” ($z < 0$) and compares 
for $z \ge 0$ to the difference $|\phi(z)|^2 - z / g$
with the Thomas-Fermi density.
This is recapped in Appendix A.2 %~\ref{a:dipole} 
of \citet{phase-shift-II}.
The other example is an approximate calculation of the phase shift 
with the WKB-approximation for $\uTilde$ 
(expected to be valid at high energies),
using the Thomas-Fermi approximation to the potential $U_{\rm ad}$.
As shown in~\ref{a:WKB-phase}, a cancellation of two large terms
leads to $\delta^{({\rm WKB})}(E) = 0$.
A final example is provided by the analysis of 
\citet{Fetter98a} for the polynomial solutions in a harmonic trap.
The boundary layers near the turning points (constructed by linearising 
the potential there) shift the eigenvalues only in the order $(\ell/R)^3$ 
where $R$ is the distance between the two turning points.
This points to a ``zero phase shift'' in the linear potential,
while the parabolic confinement in the bulk of the system provides
the main contribution to the energy eigenvalues.
We show in the following section~\ref{s:discrete-spectrum} 
how this can be phrased in the spirit of the Bohr-Sommerfeld formula.

\section{Applications}
\label{s:applications}

%Border corrections to a homogeneous system, continuous spectrum.

\subsection{Discrete spectrum}
\label{s:discrete-spectrum}

\begin{figure}[htb]
\centerline{%
\begin{minipage}[b]{0.5\textwidth}
\includegraphics*[scale=0.47]{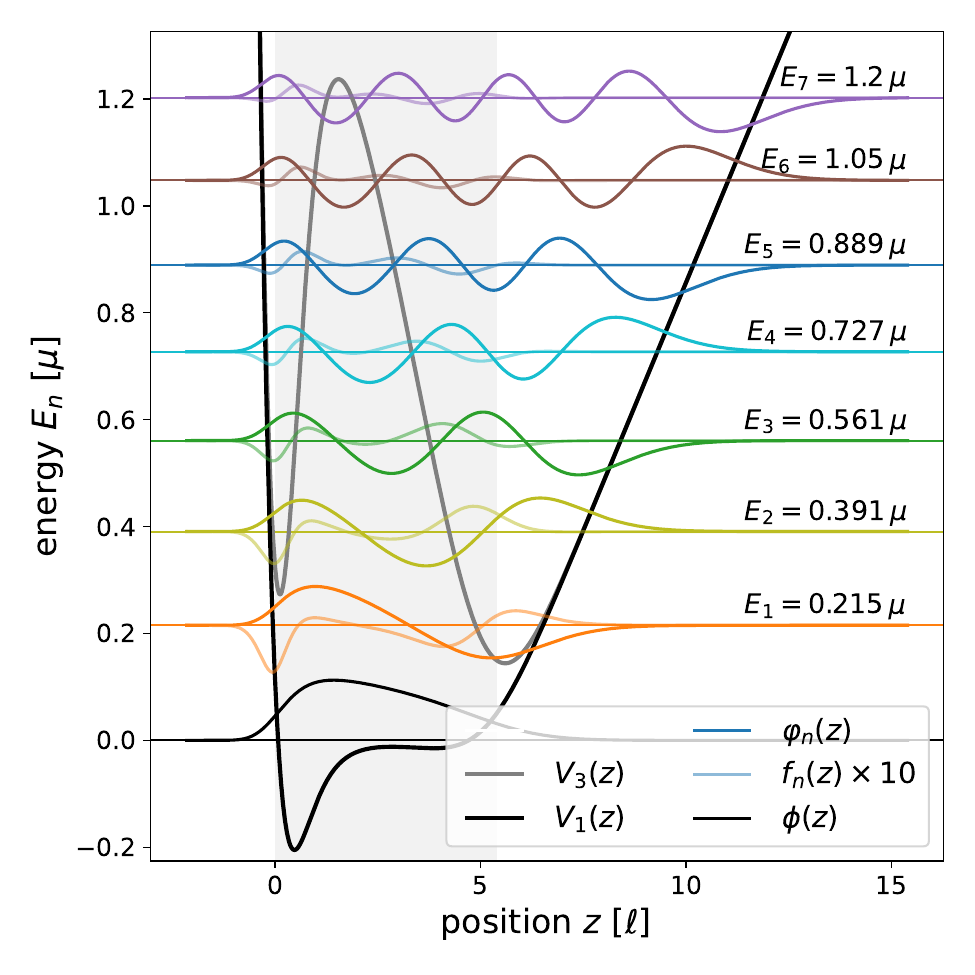}
\end{minipage}
\hspace*{0mm}
\begin{minipage}[b]{0.4\textwidth}
\includegraphics*[scale=0.42]{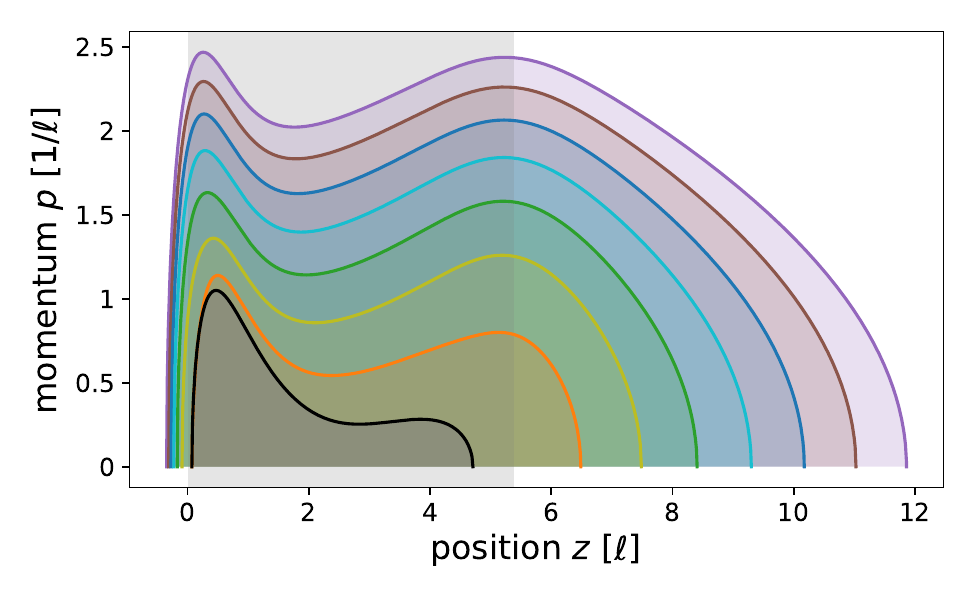}
\\[0mm]
\includegraphics*[scale=0.42]{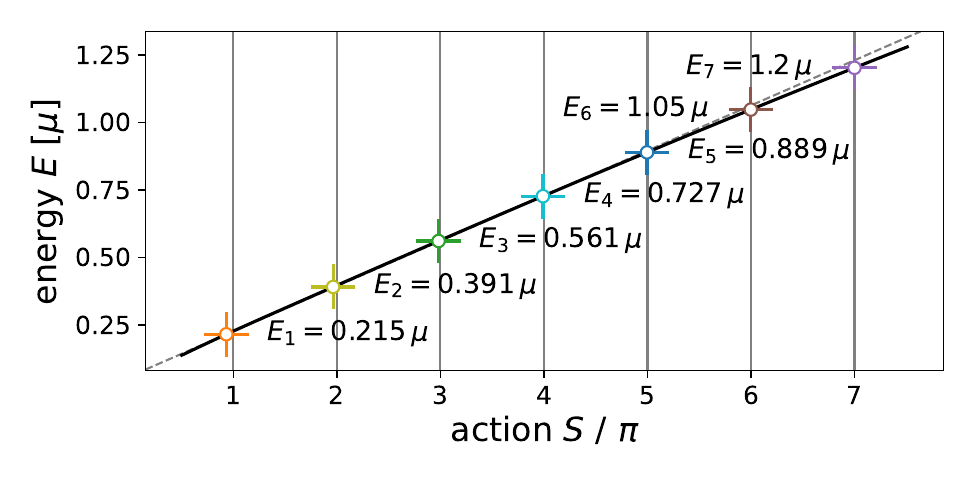}
\end{minipage}
}
\caption[]{%
%• • • 
%repeat with smaller value $1/a \approx 50\,{\rm nm}$
% ... no, we want the three length scales to be rather close
%• • •
(\emph{left}) Bogoliubov modes in a gravitational (linear)
trap closed at the “bottom” by an exponential barrier. 
The phase $\uTilde_n$ and density $\vTilde_n$ modes in the adiabatic
basis are plotted (solid and faded curves). The condensate is
the bottom black curve (not to scale).
Positions are scaled
to the characteristic length $\ell$ in the linear potential.
The gray shaded area marks the two turning points in the 
potential $V(z)$.
(\emph{top right}) Phase space portrait (only one half is shown)
corresponding to the adiabatic potentials $U_{\rm ad}(z, E_n)$
for the phase modes, evaluated at the quantised levels $E_n$.
(\emph{bottom right}) Comparison to the Bohr-Sommerfeld
quantisation rule $S = S_{12} - \pi/2 = n \pi$ 
where the action $S_{12}(E)$ is computed between the turning points 
in the potential $U_{\rm ad}(z, E)$. 
The dashed line is a linear interpolation through the low-lying levels.
Parameters are given after Eq.\,(\ref{eq:def-grav-trap}).}
\label{fig:grav-trap}
\end{figure}

As a first application of the reflection phase shifts, consider a
potential well closed by two reflecting barriers.
The Bogoliubov spectrum is then discrete, and we may expect the
phase shifts at two turning points to contribute to the total round trip 
phase (or action) in the Bohr-Sommerfeld quantisation rule
\begin{equation}
S_{12}(E) = \int_{z_1}^{z_2}\!{\rm d}z \, p(z, E)
= n \pi - \delta_1 - \delta_2
\label{eq:Bohr-Sommerfeld-phase}
\end{equation}
where $p$ is the momentum in the relevant potential.
Consider as a simple, but nontrivial example the “gravitational” trap 
with a shallow linear turning point at the “top” 
and an exponential barrier on the “bottom”
\citep{Wallis92, Henkel94a, Wilkens95b}:
\begin{equation}
V(z) = \mu\, \exp(- a z) + F z
\label{eq:def-grav-trap}
\end{equation}
For definiteness, we take $1/a \approx 130\,{\rm nm}$, 
%• • • 
%take smaller value
%• • •
gravity for $F$, 
and $\mu \approx 166\,{\rm nK}$,
with other parameters as in Table~\ref{t:units}.
The ratios of decay length, healing length, and characteristic length 
of the linear potential are then $1/a : \xi : \ell = 1 : 1.4 : 2.32$.
The resulting condensate solution is the solid black curve in the
bottom of Fig.\,\ref{fig:grav-trap}(left).

It is a challenge to identify the contribution of the phase shifts
because the length scales are not well separated, and one has to
split the total phase between the two turning points in a meaningful way
[see Eq.\,(\ref{eq:Bohr-Sommerfeld-phase})]. 
In the adiabatic approximation with the potential $U_{\rm ad}$,
we would use the momentum $p = \sqrt{- U_{\rm ad}}$ 
in Eq.\,(\ref{eq:Bohr-Sommerfeld-phase}).
We have checked that this actually reproduces (i.e., includes) 
the phase shift computed in the previous sections. 
The potential $V_1$ shown in Fig.\,\ref{fig:grav-trap}(left)
has a narrow minimum at the left (steep) turning
point and is nearly flat when approaching the right (soft) turning
point. 

We also plot the phase
and density modes $\uTilde_n$, $\vTilde_n$ as defined by 
Eqs.\,(\ref{eq:rotated-basis}, \ref{eq:def-adiabatic-angle}).
The density modes are essentially localised in the
condensate region (gray shaded area). 
For comparison, the energy-independent
potential $V_3$ is shown in solid gray.
The density modes qualitatively behave as expected from this double-well 
if the tunnelling coupling is relatively strong.
The phase modes are dominant in determining the quantised energy
levels, as shown by the comparison to the semiclassical action
computed in the adiabatic potential $U_{\rm ad}(z, E)$ 
(Fig.\,\ref{fig:grav-trap}, right column).
%The Bohr-Sommerfeld rule thus deviates only very little at low energies 
%where the semiclassical WKB technique is expected to fail anyway.

%The impact on the energy levels
%of the phase shifts $\delta(E)$ at the turning points
%is barely visible in Fig.\,\ref{fig:grav-trap}. 
The Bogoliubov levels in the gravitational trap 
are nearly equidistant (see the left panel in Fig.\,\ref{fig:grav-trap},
and the dashed diagonal in the bottom right).
This would be expected for a flat-bottom potential in the low-energy limit 
where the Bogoliubov dispersion is linear,
similar to the approximation applied by \citet{Oehberg97}.
Actually, the WKB
calculation reproduces most of the features of the spectrum, 
except at very low energies where small deviations
from the Bohr-Sommerfeld rule (see caption) are visible.
In particular,  
the lowest points in Fig.\,\ref{fig:grav-trap}(bottom right)
deviate somewhat from integer multiples of $\pi$.

\subsection{Boundary correction to bulk density of states}

A more direct way to explore the relevance of the phase shift
is to consider a flat-bottom potential so that for $z \to \infty$,
we expect the system to behave like a homogeneous one.
The following recipe to count states owes a lot to \citet{Barton79}.
We apply the same device as in the numerics
shown in Fig.\:\ref{fig:adiabatic-phif} and
enclose the system in a large, finite box.
Consider the case of a Dirichlet boundary condition
at $z = L$, larger than any other scale.
(We emphasise that this boundary condition is imposed only on the Bogoliubov modes, 
not on the condensate.) 
From the 
asymptotes~(\ref{eq:asymptotic-u-and-v-0}) of the $u$ and $v$ modes, 
%asymptotic form of the reflected waves,
%\begin{equation}
%\psi_k(z) = \frac{\sin( k z + \delta(k) )}{\sqrt{k}} 
%\label{eq:}
%\end{equation}
we get for the allowed energies $E_k$ or $k$-vectors in a flat-bottom potential
the quantisation rule
\begin{equation}
k L + \delta(k) = n \pi 
\,, \quad n = 1, 2 \ldots
\label{eq:quantisation}
\end{equation}
Consider now some quantity $Q$ to be evaluated by summing
over all modes, like the mean energy in thermal equilibrium 
$E_k \langle b_k^\dag b_k \rangle$ or the ground state energy.
As $L$ is large, the modes become dense and the sum turns into an integral
\begin{equation}
Q = \sum_k Q_k \approx \int\limits_0^\infty\!{\rm d}k\, \rho(k) \, Q(k)
\label{eq:F-sum-over-modes}
\end{equation}
where $\rho(k)$ is the density of modes in $k$-space. For the simple
reference case of a box potential between $z = 0$ and $z = L$, the
spacing of the discrete $k$-vectors is $\Delta k = \pi/L$ so that
$\rho(k) = L / \pi$. This makes $Q/L \to q$ an intensive quantity 
in the thermodynamic limit.

The phase shift $\delta(k)$ from the left boundary
changes the mode spacing according to 
[subtract Eq.\,(\ref{eq:quantisation}) evaluated for $n+1$ and $n$]
\begin{equation}
(k_{n+1} - k_{n})L + \delta(k_{n+1}) - \delta(k_{n}) = \pi
\label{eq:}
\end{equation}
The mode density 
$1/(k_{n+1} - k_{n}) = 1/\Delta k$ becomes for large $L$, after expanding
the phase shift
\begin{equation}
% L \Delta k + \frac{ \pi }{ L } \frac{ {\rm d} \delta }{ {\rm d}k } = \pi
\rho(k) = 
\frac{ 1 }{ \Delta k } = 
\frac{ L }{ \pi }
\left( 
1  - \frac{ 1 }{ L } \frac{ {\rm d} \delta }{ {\rm d}k }
\right)^{-1}
\approx 
\frac{ L }{ \pi }
+
\frac{ 1 }{ \pi } \frac{ {\rm d} \delta }{ {\rm d}k }
\label{eq:}
\end{equation}
Inserting into the mode sum~(\ref{eq:F-sum-over-modes}) and subtracting
the extensive part $\sim L$, we get the boundary shift $\Delta Q_B$ 
of the quantity $Q$ relative to its extensive limit
\begin{equation}
\Delta Q_B = \lim\limits_{L \to \infty} \left( Q(L) - L q \right) 
=
\int\!\frac{ {\rm d}k }{ \pi }  
\frac{ {\rm d} \delta }{ {\rm d}k } \, Q(k)
\label{eq:shift-in-mode-density}
\end{equation}
Note that the derivative of the phase shift plays the role of 
a change in mode density. 
This technique can be used to compute, e.g., shifts in 
zero-point energies related to the boundary conditions
\citep{Barton79}.
We do not pursue this issue further here because in the
companion paper \citep{phase-shift-II}, the quantities of
interest (like the ground-state energy density)
depend on the spatial profile of the modes,
a feature that is not captured by Eq.\,(\ref{eq:shift-in-mode-density}).

%• • • 
%Not pursued further because of more detailed description below,
%using the
%Bogoliubov modes in the inhomogeneous background.
%• • •

\begin{figure}[bht]
\centerline{%
\includegraphics*[height=0.3\textwidth]{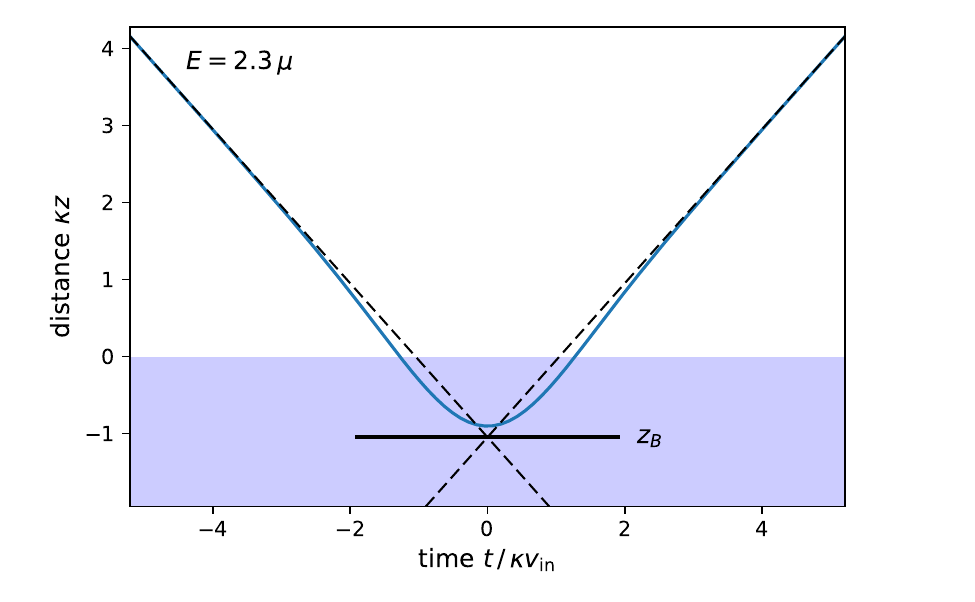}
\hspace*{02mm}
\includegraphics*[height=0.3\textwidth]{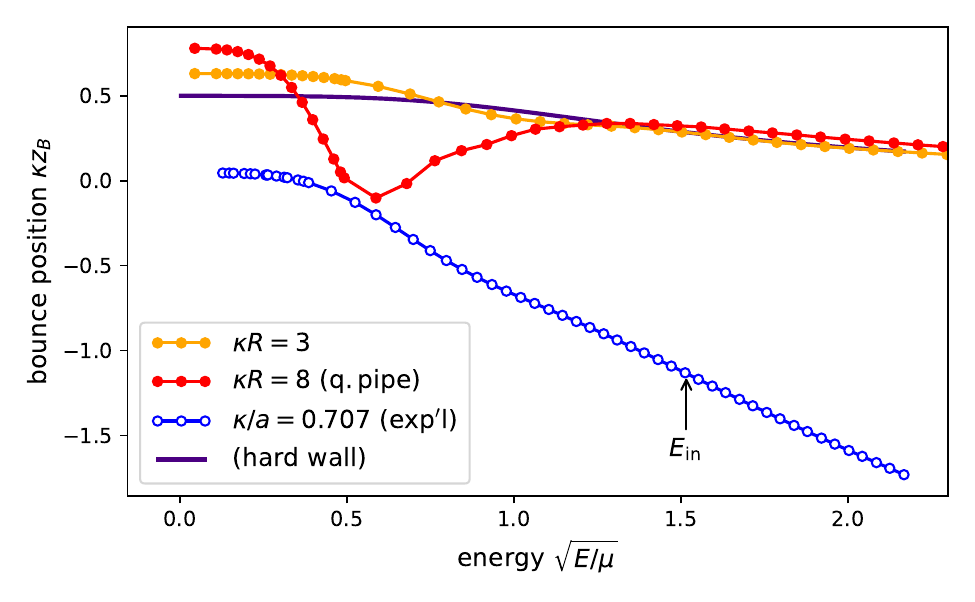}
}
\caption[]{%
(\emph{left}) 
Illustration of bouncing trajectory in the exponential
potential, $\kappa/a = 1/\sqrt{2}$. 
The two asymptotes of the trajectory (computed numerically by solving
Newton's equation of motion in the potential $U_{\rm ad}(z; E)$) 
cross at the thick horizontal line $z = z_B$.
(\emph{right}) 
Position of bounce, $z_B = - {\rm d}\delta / {\rm d}k$
for Bogoliubov wave packets reflected off the condensate edge.
Here, $k = k(E)$ is the asymptotic momentum far from the turning
point, given by the Bogoliubov dispersion relation. (Only flat-bottom
potentials are considered here.)
}
\label{fig:bounce}
\end{figure}

The same derivative as in Eq.\,(\ref{eq:shift-in-mode-density})
appears, however, when a wave packet of Bogoliubov excitations
with a well-defined mean position approaches the turning point 
and gets reflected back into the bulk.
When the trajectories of the incident and reflected wave packet are
extrapolated, they cross at $z_B = -{\rm d} \delta / {\rm d}k$, as if
the wave packet had bounced off at this position \citep{Henkel94a},
see Fig.\,\ref{fig:bounce}(left).
The phase shifts found in this paper for flat-bottom potentials yield
effective bounce positions shown in the right panel. 
Consider a particle in two or three dimensions that
bounces at the plane $z = 0$ 
where the turning point for the condensate wave function is located.
For an oblique incidence, 
the reflected trajectory is also laterally shifted,
as can be inferred from Fig.\,\ref{fig:bounce}(left).
This is the key idea behind the optical ray shift
of \citet{Goos47}.

%• • • 
%how is the force or group velocity computed? in which potential?
%for the $u$ or the $\uTilde$ wave function?
%• • •

%Change in mode density from ${\rm d} \delta / {\rm d}E$ (inverse “dwell time”),
%evaluate depleted atom number (spectrum)?

\section{Conclusion}

The reflection of waves from a potential is an archetypal problem of scattering
theory with many implications for cross sections, tunnelling delay times,
or mode counting, to mention a few. We have here analysed this case
for elementary excitations of a Bose-Einstein quasi-condensate confined
by a half-open potential, covering both hard and soft barriers and the
entire continuous spectrum. Wave reflection is dispersive (with an energy-dependent
phase shift)
even for a hard wall which is mostly due to the condensate background 
being inhomogeneous. The results are obtained with a combined numerical and
analytical scheme that removes divergent solutions from the Bogoliubov
equations. As applications, we have studied the energy spectrum of a trapped 
system and the mirror position for a bouncing wave packet. The case of a 
linear potential is special in that its phase shift is numerically relatively
small at all energies, provided the slow variations of the phase shift
in the reflected Bogoliubov wave are tracked over a sufficiently large region 
into the bulk of the trapped gas.

\paragraph{Acknowledgments.}
For calculations and discussions at various stages of this work,
I thank Uwe Bandelow, Abdoulaye Diallo, Enrico Reuß, Anja Seegebrecht,
and the late Gabriel Barton.

% SFB 1636

\appendix

\section{WKB phase shift for the linear potential}
\label{a:WKB-phase}

We discuss here an approximate calculation of the phase shift in the
linear potential and
argue that the small values we find
arise from the cancellation between two large contributions.
The idea is to solve the Schr\"odinger equation~(\ref{eq:adiabatic-for-u})
with the WKB approximation~\citep{MessiahI}
\begin{equation}
z > z_*: \qquad
\uTilde^{({\rm WKB})}( z ) =
\frac{ A' }{ [-U_{\rm ad}( z )]^{1/4} }
\sin\bigg( \!
\frac{\pi }{ 4 } +
\int_{z_*}^{z}\!{\rm d}z'\,
[-U_{\rm ad}( z' )]^{1/2}
\bigg)
	\label{eq:wkb-solution}
\end{equation}
where $p(z) = [-U_{\rm ad}( z )]^{1/2}$ is the semiclassical momentum
and $z_*$ the left turning point: $p( z_* ) = 0$.
We are only interested in the asymptotic regime $z \to \infty$
and evaluate the action (or phase) integral approximately.
This can be done analytically
when the Thomas-Fermi approximation is used (adopting natural units
for the linear potential):
\begin{equation}
- U_{\rm ad}(z) \approx
\left\{
\begin{array}{ll}
\displaystyle
E + \mu - V(z) \approx E + z
\,,
& \mbox{for } z \le 0
\,,
\\[1ex]
\displaystyle
\sqrt{ E^2 + g^2 |\phi(z)|^4 } - g |\phi(z)|^2
\approx \sqrt{ E^2 + z^2 } - z 
\,,
& \mbox{for } z \ge 0
\,.
\end{array}
\right.
\label{eq:}
\end{equation}
We thus have $z_* = -E$, and the interval $z' = - E \ldots 0$
contributes the phase
\begin{equation}
\int_{-E}^{0}\!{\rm d}z'\,
\sqrt{ E + z' }
=
\frac23 E^{3/2}
\label{eq:}
\end{equation}
Between $z' = 0 \ldots z$, the integration yields
\begin{eqnarray}
\int_{0}^{z}\!{\rm d}z'\,
( \sqrt{ E^2 + z^{\prime 2} } - z' )^{1/2}
&=& \frac{2E}{3} 
\frac{ 
\sqrt{ E^2 + z^2 } + 2 z
}{( \sqrt{ E^2 + z^2 } + z )^{1/2}} 
%\left( \sqrt{ E^2 + z^2 } + 2 z \right)
%( \sqrt{ E^2 + z^2 } - z )^{1/2}
- \frac23 E^{3/2}
\nonumber\\
&=&
E \sqrt{ 2 z } - \frac23 E^{3/2} + 
{\cal O}( z^{-3/2} )
% 	\frac{ E^3 }{ 6 \sqrt{ 2 z^3 } }
%\,,\qquad
%\mbox{for } z \gg E
\label{eq:}
\end{eqnarray}
In the last line, we expanded for $z \gg E$ to recover the
square root scaling of the phase.

Note the cancellation of the constant term $- E^{3/2}$ when the 
two contributions are summed. Coming back to the wave 
function~(\ref{eq:wkb-solution}) and comparing to the asymptotic
form~(\ref{eq:asymptotic-Bessel-Coulomb}), the phase
offset $\pi / 4$ is matched by the WKB phase
originating from the turning point $z_*$~\citep{BenderOrszag,MessiahI}.
This combination of WKB and Thomas-Fermi approximations
therefore suggests a phase shift $\delta( E ) = 0$, 
in agreement with the trend observed
in Fig.\,\ref{fig:phase-shift-1}(right) for large $E$. 
We note that 
it is only in this regime that we expect 
the WKB and the adiabatic approximations
to be accurate because the local wavelength
is small and varies slowly enough over the entire classically allowed 
region.

\section{Additional plots}
\label{a:plots}

\subsection{Illustration of adiabatic potential}

%Split $U_{\rm ad}$ in three pieces: beyond-Thomas-Fermi, 
%square root difference and geometric.

\begin{figure}[bhtp]
\centerline{%
\includegraphics*[scale=0.42]{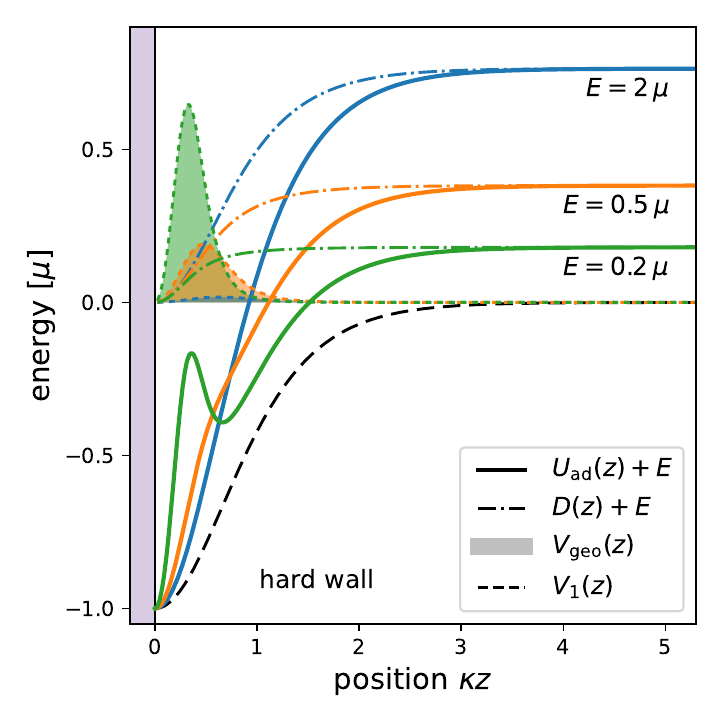}%
\hspace*{3mm}%
\includegraphics*[scale=0.42]{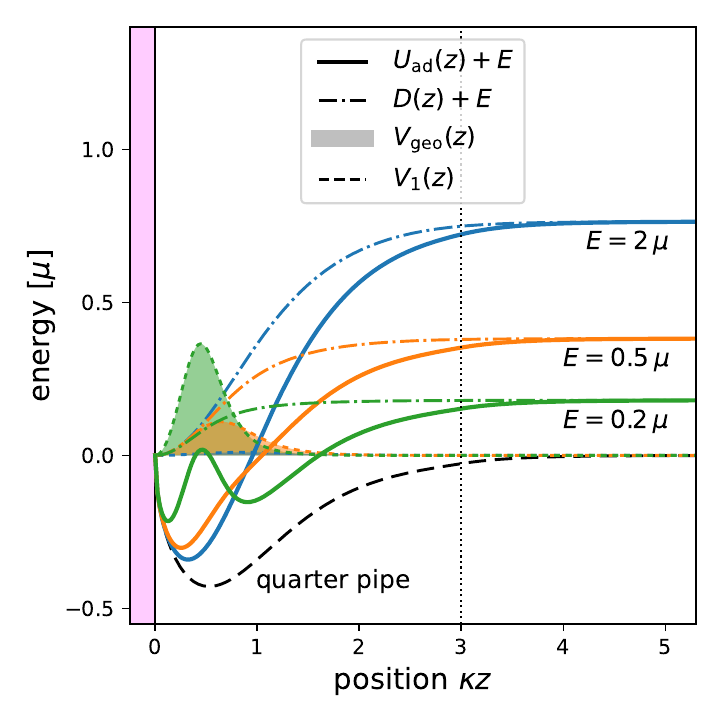}}

\centerline{%
\includegraphics*[scale=0.42]{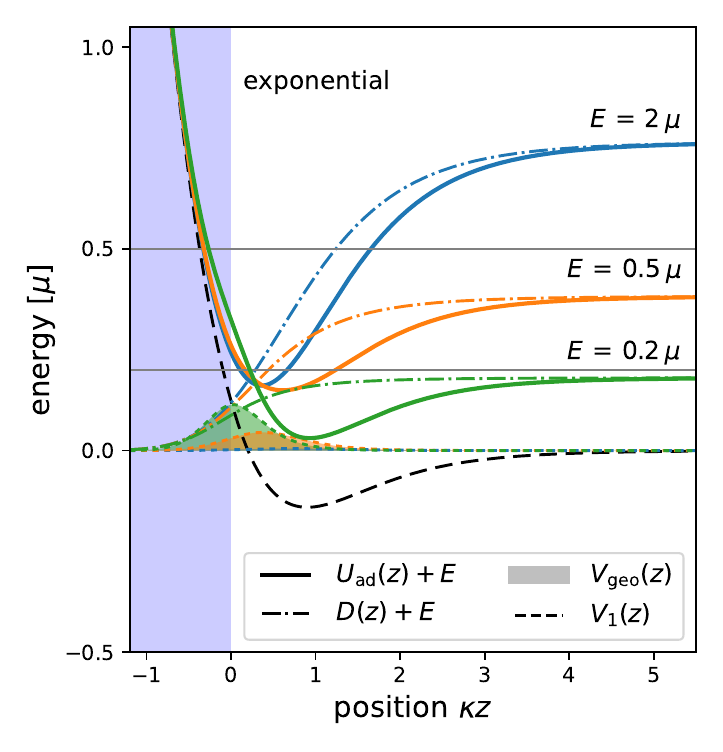}%
\hspace*{3mm}%
\includegraphics*[scale=0.42]{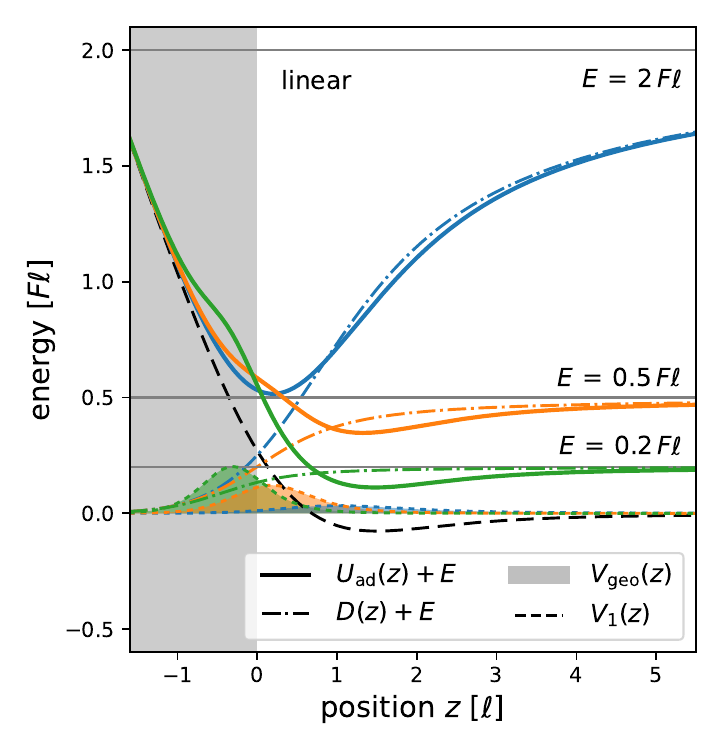}}

\caption[]{%
%• • • 
%exchange in legend $V_1$ and $V_{\rm geo}$
%• • •
%•
%set right end at $z = 5$ in top row -- to visualise the slightly
%wider potential well in the quarter pipe case.
%•
%
%• • •
%is that Figure really needed?
%• • •
Decomposition of the adiabatic potential $U_{\rm ad}$ in three
terms according to Eq.\,(\ref{eq:split_U_ad}). 
The color shaded areas give the geometric potential $V_{\rm geo}$,
it nearly vanishes for the largest chosen energy.
Numerical parameters as in Table~\ref{t:units}, 
$\kappa R = 3$ for the quarter-pipe case,
$\kappa = a$ for the exponential one.
% • quarter pipe: polish xlim, ylim
% • exponential: no $[\mu]$ in ylabel
}
\label{fig:split_U_ad}
\end{figure}

The potential $U_{\rm ad}(z; E)$ for the phase mode $\uTilde$
is illustrated in Fig.\:\ref{fig:split_U_ad}, where also the
three pieces $V_1$, $D$, and $V_{\rm geo}$ introduced in
Eq.\,(\ref{eq:split_U_ad}) are displayed separately.

\subsection{Trapped density mode}

%indeed relevant for a density fluctuation spectrum. But keep it short, 
%mainly to illustrate that at not too low energy, the adiabatic scheme
%performs really well.

\begin{figure}[t!bh]
\centerline{%
\includegraphics*[height=0.3\textwidth]{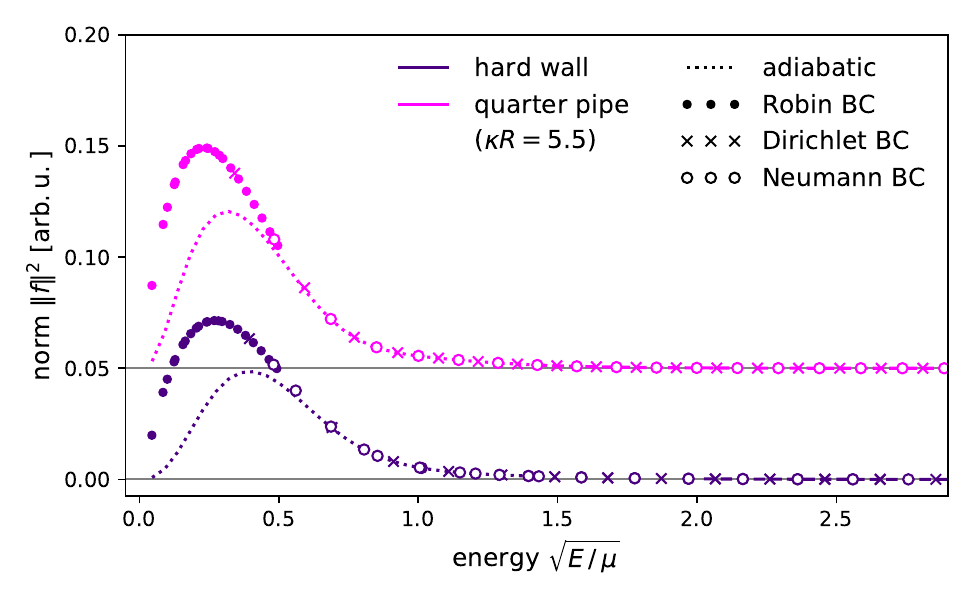}%
\hspace*{3mm}%
\includegraphics*[height=0.3\textwidth]{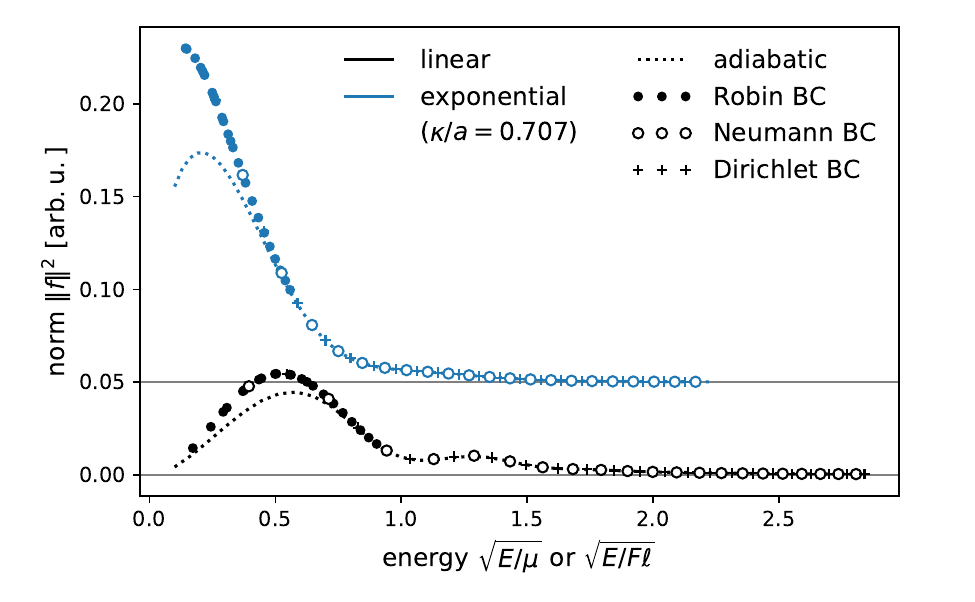}%
}
\caption[]{%
%• • • 
%write `Robin' rather than `log deriv'; 
%plot adiabatic in dotted (not dashed)
%• • •
Squared norm of the “density mode” $\vTilde(z)$ localised 
near the condensate boundary layer. 
%• • • 
%physical unit of this quantity?
%• • • 
It is plotted vs.\ the root of the mode energy $E$ for better
visibility, and some curves have been shifted vertically.
Symbols: numerical diagonalisation of the Bogoliubov equations with
boundary condition at the open right end as indicated,
dotted lines: adiabatic approximation. 
The latter solves first for $\uTilde^{({\rm ad})}(z)$ 
in Eq.\,(\ref{eq:adiabatic-for-u}), evaluates the source
term $-\hat{L}\uTilde^{({\rm ad})}(z)$ and then computes $\vTilde^{({\rm ad})}(z)$ 
by inverting the inhomogeneous
Eq.\,(\ref{eq:BdG-adiabatic-f}) on a discrete spatial grid.
Parameters as in Table~\ref{t:units},
$\kappa R = 5.5$ for the quarter-pipe case and $\mu = 1$,
$\kappa / a = 1/\sqrt{2}$ for the exponential case.
}
\label{fig:norm-v}
\end{figure}

An overview of the relative size of the $\uTilde$ and $\vTilde$ 
mode functions
is provided in Fig.\,\ref{fig:norm-v} where the squared norm
\begin{equation}
\Vert \vTilde \Vert^2 = \int\!{\rm d}x\, |\vTilde(z; E)|^2
\label{eq:}
\end{equation}
of the density mode trapped “below” the potential $V_{\rm ad}$
is shown.
The overall normalisation is such that the function
$\uTilde(z; E)$ 
behaves asymptotically according to 
Eqs.\,(\ref{eq:asymptotic-Bessel-Coulomb}, \ref{eq:def-phase-1}) with $A = 1$.
Recall that $f$ is localised in those non-adiabatic regions where the condensate 
density changes rapidly (differential operator $\hat{L}$ in
Eqs.\,(\ref{eq:BdG-adiabatic-phi}, \ref{eq:BdG-adiabatic-f}, \ref{eq:def-L})).
The values for $\Vert \vTilde \Vert^2$ are relatively small 
and smoothly vanish as the energy becomes large. 
%• • • 
%Discuss low-$E$ behaviour
%• • •
For all potentials, a maximum is observed at low energies whose
value is largest for the exponential potential. 
(The low energies shown as full dots are 
reached numerically using a Robin boundary
condition intermediate between Dirichlet and Neumann, 
as explained in Sec.\:\ref{s:numerics}.)
%Note the difference to the linear potential where 
%$\Vert \vTilde \Vert^2$ goes back to zero as $E \to 0$.
%We attribute this to the normalisation of $\uTilde$: 
%it is dominated by the large-$z$ regime which is quite different
%for a flat-bottom and the linear potential [compare
%Eqs.\,(\ref{eq:def-phase-1}, \ref{eq:asymptotic-Bessel-Coulomb})].
For all four types of potentials, a good performance of the adiabatic
approximation is found, 
unless the energy is much below $\mu$.
Since in this regime the non-adiabatic coupling $\theta'$ is
largest in amplitude, the density mode $\vTilde$ is most significant there.

The broad oscillatory features that appear for the linear potential have been discussed by \citet{Diallo15a}. 
They arise from the anti-nodes of the source term $\hat{L} \uTilde$
that are moving across the minimum of the potential $V_{\rm ad}$
as a function of energy.

\subsection{Phase shift for quarter-pipe potential}

\begin{figure}[htbp] %  figure placement: here, top, bottom, or page
   \centering
   \includegraphics[height=0.29\textwidth]{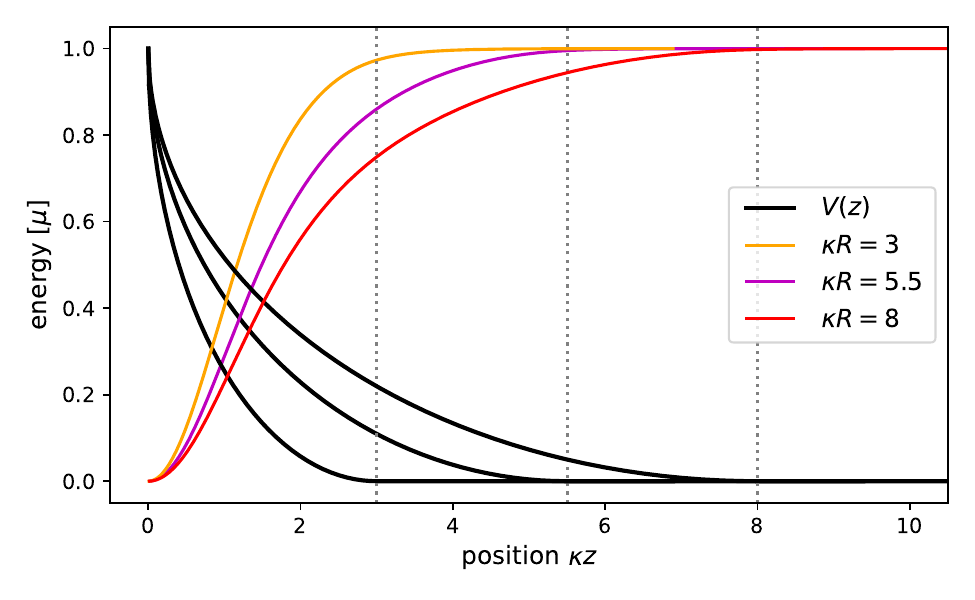}
   \hspace*{04mm}
   \includegraphics[height=0.29\textwidth]{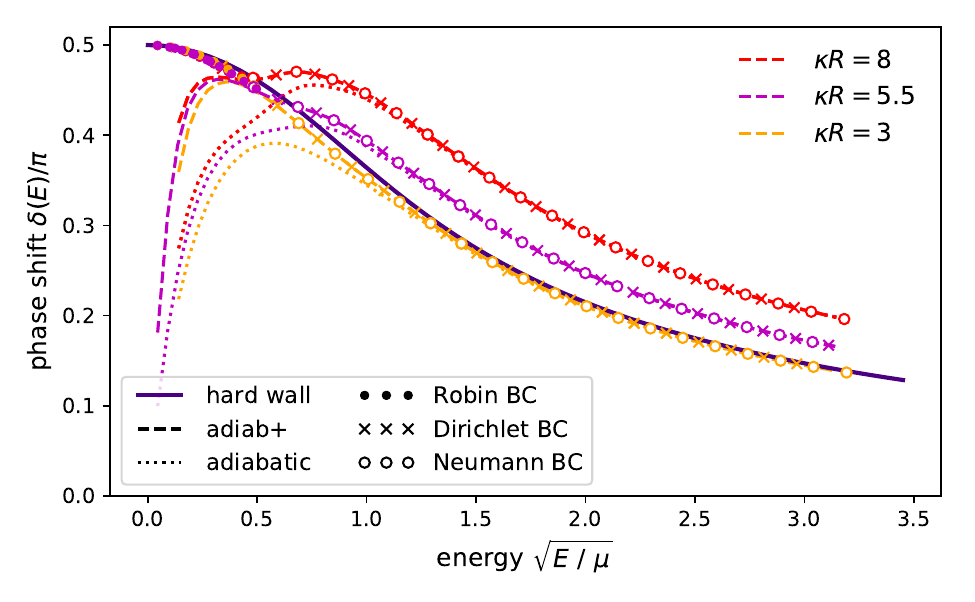}
   \caption[]{(\emph{left}) Quarter-pipe potential (black) 
   and condensate density
   $g |\phi(z)|^2$ (coloured) 
   for different values of the “pipe radius” $R$ 
   (lengths scaled to the healing length $1/\kappa$).
   (\emph{right}) Corresponding phase shift.
   Symbols: numerical solution of the Bogoliubov equations;
   dashed: adiabatic approximation including the correction
   due to the density mode [Sec.\,\ref{s:phase-shift-2},
   Eq.\,(\ref{eq:phase-shift-like-Feshbach})];
   dotted: without this correction.
   Other parameters as in Fig.\,\ref{fig:phase-shift-1}.
%   • • • 
%   do similar thing with the scale parameter $a$ for the exponential
%   potential 
%   • • •
   }
   \label{fig:scan-R}
\end{figure}

In Fig.\:\ref{fig:scan-R}, the impact of the length parameter $R$
of the quarter-pipe potential (see Table\:\ref{t:units})
is illustrated. 
On the left, the potential and its condensate density,
on the right, the phase shift as a function of energy.
The case of a wide pipe, where the deviation from the asymptotic density
covers the widest spatial range, correlates with the largest
phase shifts.
An exception are very low energies where apparently the detailed
shape of the potential becomes irrelevant, and the phase shift
behaves the same as for the hard wall potential (where formally, 
$R = 0$).

\clearpage

%\bibliographystyle{sn-mathphys-ay}% elsarticle-harv}
%\newcommand{\mybibpath}{/Users/carstenh/Work/BoxUP/documents/Biblio/}
%\bibliography{\mybibpath journals,%
%../phase_shift}

% \ifx \doiurl  \undefined \def \doiurl#1{\href{https://doi.org/#1}{DOI:\texttt{#1}}}\fi


\end{document}